\documentclass{mscs}

\usepackage{graphicx}

\usepackage[round,colon,authoryear]{natbib}
\usepackage{bm}                 
\usepackage{amsmath}            
\usepackage{amsfonts}
\usepackage{amssymb}

\newcommand{\ket}[1]{\mbox{$|#1\rangle$}}
\newcommand{\bra}[1]{\mbox{$\langle#1|$}}
\newcommand{\braket}[2]{\mbox{$\langle#1|#2\rangle$}}
\newcommand{\identity}{\leavevmode\hbox{\small1\kern-3.2pt\normalsize1}}

\title[Decoherence in quantum walks]{Decoherence in quantum walks - a review}
\author[V.~Kendon]{V\ls I\ls V\ns K\ls E\ls N\ls D\ls O\ls N
\thanks{VK is funded by a Royal Society University Research Fellowship.}\\
School of Physics and Astronomy, University of Leeds, Leeds, LS2 9JT, UK\\
Submitted 1 June 2006}

\begin{document}
\maketitle

\begin{abstract}
The development of quantum walks in the context of quantum computation,
as generalisations of random walk techniques, led
rapidly to several new quantum algorithms.  These all follow
unitary quantum evolution, apart from the final measurement.
Since logical qubits in a quantum computer must be protected
from decoherence by error correction, there is no need to 
consider decoherence at the level of algorithms.
Nonetheless, enlarging the range of quantum dynamics to include
non-unitary evolution provides a wider range of possibilities for
tuning the properties of quantum walks.  For example, small amounts
of decoherence in a quantum walk on the line can produce more
uniform spreading (a top-hat distribution), without losing the quantum
speed up.
This paper reviews the work on decoherence, and more generally on
non-unitary evolution, in quantum walks and suggests what
future questions might prove interesting to pursue in this area.
\end{abstract}

\tableofcontents

\section{Introduction}
\label{sec:intro}

The study of quantum versions of random walks is easily motivated.
They have provided the engine for several new quantum algorithms,
they are of mathematical interest in their own right,
and in physical systems they form a simple example of
coherent quantum control over single atoms or photons.
They have also found applications from the early days of
quantum computation as the clock mechanism in a Feynman computer
\citep{feynman86a}.
The role of decoherence requires more explanation.
In physical systems we must of course
consider the effects of decoherence when designing
our experiments, but this alone does not qualify it as a subject
for more general study.
Decoherence plays a fundamental role in transforming from the
quantum to the classical regime.  
Quantum walks are simple systems in which the intricacies of
decoherence can be studied both analytically and numerically,
making connections with fields such as quantum chaos.
Taking a broad view of decoherence as any process that tends to
reduce quantum coherence, we find that it provides a method for
tuning quantum random walks to improve their algorithmic properties.
Instead of considering only pure quantum dynamics, we can include
non-unitary operations in our quantum random walk algorithms,
thereby enlarging the toolbox for controlling their behaviour.

The origins of quantum walks can be traced back to the dynamics of
quantum diffusion, well-studied in the physics literature, see
for example, \cite{feynman64}, mostly as models for
physical particles moving on regular lattices.
Studies of quantum dynamics using a discrete time step described as
``quantum random walks'' appeared from the late
1980s from \cite{gudder88}, \cite{grossing88a}, and \cite{aharonov92a},
the latter with quantum optical applications in mind.
\cite{meyer96a} studied similar systems with the aim of
defining quantum cellular automata \citep{meyer96c}.  The current surge
of interest in the context of quantum information
started with \cite{farhi98a} studying continuous-time
quantum walks, and continued with \cite{aharonov00a}, \cite{ambainis01a}
and \cite{nayak00a} studying discrete-time quantum walks,
all with the goal of applying them to quantum algorithms.
As with classical random walks, there are a wide range of possible
dynamics that fit within the general concept, for example,
\cite{gottlieb03a, gottlieb04a} defined a discrete-time walk with a 
continuous space and a coin with a continuous set of states.

In this review we will start with a short overview of purely
quantum versions of random walks as studied by the quantum
information community in their quest for new quantum algorithms,
then in later sections consider the added effects of decoherence.
We will cover first discrete-time, coined quantum walks
in \S\ref{sec:qrwdisc},
then continuous-time quantum walks in \S\ref{sec:qrwcont}.
We mention algorithmic applications in \S\ref{sec:qrwalgo},
briefly describing two of the known algorithms using quantum walks.
In \S\ref{sec:qrwdiscdec} and \S\ref{sec:qrwcontdec}
we consider the effects of decoherence
in quantum walks, focusing mainly on its use as a tool for
enhancing computational speed up.
There are several proposals for implementing a quantum walk
directly in a physical system, and most of these studies also
contribute analysis of the decoherence relevant to the physical
system in question: we provide a short summary of
these studies in \S\ref{sec:qrwphys}.
In general, we deal only with quantum walks on undirected graphs, 
with a brief summary of what is known about the largely unexplored realm of
directed graphs in \S\ref{sec:directed}.

This review will assume a basic familiarity with quantum mechanics
and quantum information theory.  Those wishing to get up to speed
on these areas in the context of quantum walks are referred to
the excellent and comprehensive introductory review 
from \cite{kempe03a}, which assumes
only a basic knowledge of quantum mechanics.
Those desiring a gentle overview of quantum walks in the context
of quantum computing are referred to \cite{kendon06a}.  A brief but 
accessible review of quantum walk algorithms from a computer science
perspective (no mention of decoherence) may be found in \cite{ambainis04a}.

The mathematical results and proofs for quantum 
walks without decoherence have been presented
in great detail and elegance in the
original literature: this review summarises what is known
without presenting formal proofs but with ample references
for those interested to pursue them further.
For decoherence effects fewer analytic results have been derived:
those that exist are presented in more detail, along with summaries of
many related ideas, highlighting
the myriad of open avenues that remain to be explored.

\section{Coined (discrete-time) quantum walks}
\label{sec:qrwdisc}

In this section we describe quantum walks taking place in a discrete
space of positions, with an evolution using discrete time steps.
Just as there are many ways to express the dynamics of classical
random walks, and many variations on the basic ``drunkard's walk'',
the plethora of different studies of quantum walks have proliferated
an equally varied set of notations.  
For this review, we have chosen one of the more commonly used approaches
and we try to maintain a consistent notation, with mention
of alternative methods where appropriate.
To keep the notation simple, we have taken the liberty of using the
same symbols for common quantities throughout, with the precise definition
implied by the context.  For example, $P(x,t)$ is a probability
distribution over a set of positions denoted by $x$ at a time $t$ during 
the evolution.  In the section on the quantum walk on the line,
$P(x,t)$ is for the walk on the line, in the section for the walk on a
cycle $P(x,t)$ is for the walk on a cycle, and so on.
Extra dependent variable may appear where needed: $P(x,a,t)$ is a
probability distribution with the coin state $a$ also specified.

We begin with a brief description of one of the simplest examples,
the quantum walk on the line.  This will serve both to set up our
notation and to indicate the methods of solution.
It will also illustrate the first key difference between quantum
and classical random walks, the enhanced rate of spreading.
This is followed by a discussion of quantum walks on cycles,
illustrating several more key differences between classical
and quantum random walks.
We will finish this section with a brief review of coined quantum walks
in higher dimensions, on general graphs and regular lattices.

\subsection{Coined quantum walk on an infinite line}
\label{ssec:2dline}

In much the same way as we now know almost everything about the
properties and possible states of two qubits -- though quantum computers
will clearly need far more than two qubits to be useful -- the simple
quantum walk on a line has now been thoroughly studied
\citep[e.g., see][]{ambainis01a,bach02a,yamasaki02a,kendon02c,brun02b,brun02c,konno02a,konno02b,carteret03a},
though there is no suggestion that it will lead to useful quantum walk
algorithms by itself.

First, recall the classical random walk dynamics we are
looking for a quantum counterpart of.  The walker starts
at the origin of an infinite line of discrete points
(labelled $\dots {-2}, {-1}, 0, 1, 2\dots$) and tosses an unbiased, two-sided
coin.  If the coin lands ``heads'' the walker makes one step in the 
positive direction; if it lands ``tails'' the walker steps in
the negative direction.  This process is repeated $T$ times, and
the position of the walker is noted: $-T\le x\le T$.  If the random walk is
repeated many times, the probability distribution obtained $P(x,T)$ is
binomial, as is well known and easily shown.  The standard
deviation of $P(x,T)$ is $\sqrt{T}$, i.e., the walker is found on 
average $\sqrt{T}$ steps from the origin after $T$ steps of the
random walk.

One obvious approach to creating a quantum counterpart of a classical
random walk is to have the quantum walker follow all possible
classical random walks in superposition.  This is not feasible in
the discrete-time quantum walk, as was shown by \cite{meyer96a},
because it is not reversible, and therefore not unitary
(which all pure quantum dynamics is required to be).
We thus make the quantum walk dynamics as similar as possible
to the classical random walk in a different way, by using
a quantum coin.  Historically, the role of the quantum coin has caused
much confusion, and an equivalent formulation presented by
\cite{watrous98a} using directed graphs seems to be preferred by some
computer scientists.  For brevity and clarity, in this review
we will stick to the formulation using a quantum coin, with brief mention of
the Watrous variant in \S\ref{sec:directed}.
The quantum walker on the line starts at the
origin and tosses a two-state quantum system (a qubit).  The
qubit coin -- also called ``chirality'' with the two states
labelled ``left'' and ''right'' in many papers --
can be in a superposition of both states, say
\begin{equation}
\ket{\psi(t=0)} = \alpha\ket{0,{-1}} + \beta e^{i\phi}\ket{0,{+1}},
\label{eq:initial}
\end{equation}
where the ``ket'' $\ket{\cdot}$ is a standard quantum notation to
indicate the complex vector of a pure quantum state.
On the r.h.s.~the kets are basis states: the first entry is the position
on the line ($x=0$ in this case),
and the second is the state of the quantum coin, where we choose to
label the two states $\ket{{\pm 1}}$.
There is an arbitrary phase $0\le\phi < 2\pi$,
while $\alpha$ and $\beta$ are real numbers, with
normalisation $\alpha^2+\beta^2=1$.
The quantum walker then steps in both directions according to
the state of the quantum coin, giving
\begin{equation}
\ket{\psi(t=1)} = \alpha\ket{{-1},{-1}} + \beta e^{i\phi}\ket{1,{+1}}.
\label{eq:shift}
\end{equation}
Keeping to a pure quantum (i.e., unitary) evolution for the moment,
the coin toss and step can be written as unitary operators $\mathbf{C}$
and $\mathbf{S}$.  A single step of the quantum random walk is thus
\begin{equation}
\ket{\psi(t+1)} = \mathbf{S}\mathbf{C}\ket{\psi(t)},
\end{equation}
where $\ket{\psi(t)}$ can be expanded as a superposition of basis states,
\begin{equation}
\ket{\psi(t)} = \sum_{x,c} a_{x,c}(t)\ket{x,c},
\label{eq:basis}
\end{equation}
with complex coefficients $a_{x,c}(t)$
satisfying $\sum_{x,c}|a_{x,c}(t)|^2 = 1$.
After $T$ steps we have
\begin{equation}
\ket{\psi(T)} = (\mathbf{S}\mathbf{C})^T\ket{\psi(0)}.
\label{eq:qwdT}
\end{equation}

The shift operator $\mathbf{S}$ has already been specified
implicitly by equations (\ref{eq:initial}) and (\ref{eq:shift}),
it can be defined by its operation on a basis state $\ket{x,c}$,
\begin{equation}
\mathbf{S}\ket{x,c} = \ket{x+c,c}.
\label{eq:Sdef}
\end{equation}
It remains to specify the coin toss $\mathbf{C}$, which can in principle
be any unitary operator on the space of a qubit.
The equivalence up to a bias (probability $\eta$, $0\le\eta\le 1$, to
move in the positive direction, $(1-\eta)$ in the
negative direction) of all coin operators in the walk on a line
has been noted by several authors \citep{ambainis01a,bach02a,yamasaki02a}.
Expressing the unitary operator as a matrix we have
\begin{equation}
\mathbf{C}_2=\left( \begin{array}{rr}
        \sqrt{\eta} & \sqrt{1-\eta} \\
        \sqrt{1-\eta} & -\sqrt{\eta} \\
        \end{array} \right),
\label{eq:cop}
\end{equation}
is thus the only possible type of coin for the quantum walk on a line.
For $\eta=1/2$ (unbiased), $\mathbf{C}_2^{(\text{H})}$ is commonly known
as a Hadamard operator,
\begin{equation}
\mathbf{C}_2^{(\text{H})}=\frac{1}{\sqrt{2}}\left( \begin{array}{rr}
        1 & 1 \\
        1 & -1 \\
        \end{array} \right),
\label{eq:had}
\end{equation}
thus the simple quantum walk on a line is also known as a Hadamard walk.
Trivial cases $\eta=0$,1 give oscillatory motion and uniform motion 
respectively.
The full range of behaviour is obtained by choosing different
initial coin states, i.e., varying $\alpha$, $\beta$, $\phi$
in equation~(\ref{eq:initial}).

The quantum walk on the line can be solved analytically in many
ways.  Straightforward generalisations of methods
that work well for classical random walks, such as path counting,
and Fourier transformation were both used by \cite{ambainis01a}.
We will give examples using these methods in \S\ref{sec:qrwdiscdec}.
Path counting (path integrals) was further refined in \cite{carteret03a},
and a third method using the algebra of the matrix
operators was presented by \cite{konno02b,konno02c}.  Solution using
the tools of classical (wave) optics can be found in
\cite{knight03a,knight03c}.
\cite{romanelli03a} analyse the walk on the line by separating the dynamics
into Markovian and interference terms, allowing an alternative method
of solution for the long-time limits, which they relate to the dynamics
of a kicked rotor.  They also show that in the continuum limit one obtains
the diffusion equation with added interference terms.

\begin{figure}
  \begin{center}
    \resizebox{0.6\textwidth}{!}{\includegraphics{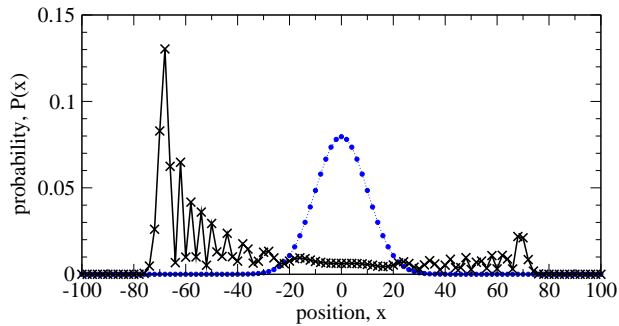}}
  \end{center}
  \caption{Comparison of a classical (dots) random walk and
	a quantum (crosses) walk with initial coin state $\ket{{-1}}$
	and a Hadamard coin operator on a line after 100 steps.
	Only even points are shown since odd points are unoccupied at even
	time steps (and vice versa).}
  \label{fig:qvscwalk}
\end{figure}
The best way to appreciate the most interesting properties
of the solution is in a graph, shown in figure~\ref{fig:qvscwalk},
of the probability distribution of the position of the walker,
\begin{equation}
P(x,T) = \sum_{c=-1}^{+1} |\braket{c}{\psi(T)}|^2,
\end{equation}
obtained in the standard way by tracing out the coin and taking the
square modulus of the wavefunction.
The quantum walk looks nothing like the classical random walk, it
spreads out much faster in a spiky distribution that is a discrete
form of an Airey function \citep{carteret03a}.
It is also asymmetric, with the asymmetry determined by the initial
coin state.  A symmetric distribution can be obtained by choosing
the initial coin state as either $(\ket{{-1}} \pm i\ket{{+1}})/\sqrt{2}$,
or $\cos(\pi/8)\ket{{-1}} - \sin(\pi/8)\ket{{+1}}$, see
\cite{konno02a,tregenna03a}.
The moments have been calculated by \cite{ambainis01a},
for asymptotically large times $T$ for a walk starting at the origin,
\begin{equation}
\sigma^2(T) = \langle x^2\rangle = (1-1/\sqrt{2})T^2,
\label{eq:sigma2}
\end{equation}
independent of initial coin state, and
\begin{equation}\label{depmoments}
\langle x \rangle_a = a(1-1/\sqrt{2})T,
\end{equation}
where $a\in\{1,-1\}$ is the initial coin state.
The standard deviation (from the origin) $\sigma(T)$
is thus linear in $T$, in contrast to $\sqrt{T}$ for the classical walk.
Moreover, unlike the classical random walk, the quantum walk evolution
depends on the initial state at all subsequent times.

\subsection{Coined quantum walk on a $N$-cycle: mixing times}
\label{ssec:2dcyc}

The $N$-cycle is the Cayley graph of the cyclic group of size $N$.
It is also a line segment with periodic boundary conditions applied,
so there is little extra work to do to apply the solution
of the quantum walk on the line to the quantum walk on the cycle.
In general, when a quantum walk occurs on the Cayley graph of some
group, it simplifies greatly on consideration of the
Fourier space of the particle \citep{aharonov00a}.
On cycles we are interested not in how far the walker strays from its
starting point, but in the mixing properties of the distribution
on the cycle.  Classical random walks on a cycle mix to a uniform
distribution in a time $O(N^2)$.  That is, after $O(N^2)$ time steps,
the walk has all but forgotten its starting state and is equally likely
to be found on any site around the cycle.  More formally, we can 
choose any small $\epsilon$ and find the mixing time $M(\epsilon)$ such that
\begin{equation}
M(\epsilon)=\min\left\{T\enspace|\enspace\forall\enspace t>T: ||P(x,t)-P_u||_{\text{tv}}<\epsilon\right\}
\label{eq:mixdef}
\end{equation}
where $P_u$ is the limiting  distribution over the cycle, and
the total variational distance is defined as
\begin{equation}
||P(x,T) - P_u||_{\text{tv}} \equiv \sum_x|P(x,T) - P_u|.
\label{eq:tvd}
\end{equation}
In general, any reasonable distance function will do the job here, but
we will stick with one commonly used in the literature so we can make
quantitative comparisons.
For a classical random walk on a cycle,
$M(\epsilon) \sim O(N^2\log (1/\epsilon))$.

In the quantum walk on the cycle, the first observation is discouraging:
quantum random walks are deterministic, they oscillate forever and
in general do not mix even instantaneously.
But by defining a time-averaged distribution,
\begin{equation}
\overline{P(x,T)}=\frac{1}{T}\sum_{t=1}^{T} P(x,t),
\label{eq:meanP}
\end{equation}
quantum walks on cycles do mix.
Operationally, this just means choosing randomly a value of $t$ between $1$
and $T$, then measuring the position of the quantum walker after $t$ steps.
However, unlike the classical random walk, the limiting distribution
$\lim\{T\to\infty\}(\overline{P(x,T)})$ for a quantum walk is not, in general,
the uniform distribution.
This is in stark contrast with a classical random walk, which always mixes to
the uniform distribution (on a regular undirected graph).

Exactly as for $M(\epsilon)$ above
we can then define the mixing time for $\overline{P(x,T)}$,
which we will denote $\overline{M}(\epsilon)$,
\begin{equation}
\overline{M}(\epsilon)=\min\left\{T\enspace|\enspace\forall\enspace t>T: ||\overline{P(x,t)}-P_u||_{\text{tv}}<\epsilon\right\}.
\label{eq:meanmixdef}
\end{equation}
\cite{aharonov00a} proved that the coined quantum walk on a cycle
with an odd number of nodes does mix to the uniform distribution, and 
has $\overline{M}(\epsilon)$ bounded above by $O(\epsilon^{-3}N\log N)$,
almost quadratically faster (in $N$) than a classical random walk.
They also proved a lower bound on the time-averaged mixing times for
quantum walks on general graphs of bounded degree,
suggesting a quadratic improvement
over classical random walks is the best that can be achieved.

Notice that there is a price to be paid for the time-averaging:
the scaling of the mixing time $\overline{M}(\epsilon)$ depends on
$1/\epsilon$, compared to $\log(1/\epsilon)$ for a
classical random walk (where no time-averaging of $P(x,T)$ is needed).
\cite{aharonov00a} show that this can be avoided by including
an amplification step.  The quantum walk is run several times, each time
starting from the final state of the previous walk.  Applied in the
optimal way, their bound on the quantum mixing time reduces
to $O(N\log N \log(1/\epsilon))$.  Recent work by \cite{richter06a,richter06b}
improves this result to $O(N\log(1/\epsilon))$.  Since the intermediate
measurements render the overall quantum walk dynamics non-unitary,
we will postpone more detailed discussion to \S\ref{ssec:cycdec}.

This speed up in the rate of mixing of quantum walks is a second key
difference between quantum and classical random walks.
We can also say a little more about their non-classical limiting distributions.
For example, Hadamard walks on cycles with an odd number of
nodes converge to the uniform distribution
\citep[as was proved by ][]{aharonov00a}, but those with an even
number converge to a non-uniform distribution unless an extra phase
$\delta$ is added to the Hadamard coin operator 
\begin{equation}
\mathbf{C}_2=\left( \begin{array}{rr}
        \sqrt{\eta} & e^{i\delta}\sqrt{1-\eta} \\
        e^{-i\delta}\sqrt{1-\eta} & -\sqrt{\eta} \\
        \end{array} \right),
\label{eq:copdelta}
\end{equation}
as \cite{tregenna03a} and \cite{bednarska03a} show.
By appropriate choice of
coin operator, a walk on any size cycle can be made to converge either
to a uniform or to a non-uniform probability distribution.
In classical random walks, the properties
of the limiting distribution depend solely on the form of the graph.

\subsection{Periodicity in coined quantum walks on cycles}
\label{ssec:pd}

A further curious property of quantum walks on cycles is that
for a few special choices of cycle size $N$, a perfectly periodic
walk arises where the quantum walk returns exactly to its starting
state after a fixed number of steps $\Omega$, and then repeats
the sequence over again, returning at $2\Omega$, etc..
The classical random walk has no such behaviour,
and returns to its initial state only after irregular numbers of steps.
This periodicity is not connected with whether the
limiting distribution is uniform or not, since here we are concerned with
exact return to the initial state, rather than the time-averaged
quantity in equation~(\ref{eq:meanP}).  Some of these periodic walks
also mix instantaneously, if we allow a walk on an even cycle to 
be considered mixed on just the odd or just the even-numbered sites
(the same issue arises when considering classical random walks on
even-sized cycles).

Using a Hadamard coin, the ``cycle'' of size $N=2$ is trivially periodic,
returning to its original state after two steps.  A cycle of size $N=4$ has
a period of eight steps.  This was first noted by \cite{travaglione01a}.
\cite{tregenna03a} continued this investigation, and found
the cycle with $N=8$ has a period of 24 steps, but $N=16$ is chaotic
and does not return to its initial state exactly even after many
thousands of steps.
These periodic cycles also exhibit instantaneous mixing half way through 
on their way to returning to the initial state, but instantaneous mixing
has not been systematically studied in discrete walks on the cycle.
If the coin is allowed to be biased, a few more periodic examples
can be found, $N=6$ with period 12, and $N=10$ with period 60.
With judicious choice of phases in the coin operator,
$N=3$ has a period of 12, and $N=5$ has a period of 60, clearly
related to $N=6$ and $N=10$ respectively,
but these were the only periodic odd-$N$ cycles found.
It is not known whether 
all periodic quantum walks on a cycle have been identified,
but \cite{tregenna03a}
conjecture that there are only a finite number of such solutions
and that this is nearly all if not all of them.
Neither periodicity nor instantaneous mixing has been explored
on more general graphs beyond the 
trivial extensions of the above to the torus and twisted torus in 
\cite{tregenna03a}.

\subsection{Coined quantum walks on higher dimensional graphs}
\label{ssec:higherD}

Classical random walks are not limited to one-dimensional
structures, and neither are quantum walks.
All that is required
for the discrete-time quantum walk is a coin that is large
enough to handle the number of choices at each vertex the
quantum walker might land on.

Consider a general graph $G$, with $N$ vertices in a set $V$,
connected by edges from the set $E$.  How one proceeds depends 
on what prior knowledge one has about $V$ and $E$.  
Since we are discussing only undirected graphs, 
if vertices $x,y\in V$ are connected by an edge $e_{xy}\in E$,
then also $e_{yx}\in E$ and
one is allowed to travel both from $x$ to $y$ and from $y$ to $x$.
The usual way to represent the structure of the graph is in an
adjacency matrix $\mathbf{A}$, which has unit entries for
each $A_{xy}$ for which $e_{xy}\in E$, and zeros everywhere else.
Since $A_{xy} = A_{yx}$ for an undirected graph, $\mathbf{A}$ is symmetric.
This representation of the graph assumes that $e_{xy}$ is unique,
i.e., there is at most one edge between any two vertices in $G$.
Classically, one can subsume multiple edges between the same vertices
into a set of ``edge weights''.  However, a quantum walker might
traverse both edges at once but with different phases, which would 
in general have a different outcome to one weighted edge.

Given no further information about $G$, there could be as many
as $N(N-1)/2$ edges, corresponding to the complete graph, for
which every vertex has $N-1$ edges leading from it.  In this case
the coined quantum walk
needs to use a coin of at least this size ($N-1$), the 
details of how to implement this quantum walk may be found in
\cite{kendon03c}.  
If one knows the maximum degree of the graph, $d$,
then one only needs a coin of size $d$.  The details
of how to implement this quantum walk, originally from
\cite{watrous98a},
are more accessibly described in \cite{ambainis03a}
and \cite{kendon04a}.
A different approach, using self-loops to make the degree of
the graph constant, is mentioned in \cite{kempe03a}.

Here we summarise the approach given in \cite{kendon04a}.
First we define our Hilbert space for the quantum walk,
$\mathcal{H}_\text{vc}$
which contains the $N$-dimensional Hilbert space
\begin{equation}
        \mathcal{H}_\text{v}=\text{span}\{| x\rangle_\text{v}:x\in\mathbb{Z}_N,
                \!_\text{v}\langle x|x' \rangle_\text{v}=\delta_{xx'}\}
                \subset \mathcal{H}_\text{vc}
\label{eq:HVspan}
\end{equation}
of vertex states,
and for the coin a $d$-dimensional Hilbert space
\begin{equation}
        \mathcal{H}_\text{c}=\text{span}\{|c\rangle_\text{c}:c\in\mathbb{Z}_d\}
        \;\;\text{and}\;\;      \!_\text{c}\langle c|c' 
\rangle_\text{c}=\delta_{cc'}\}
\label{eq:HCspan}
\end{equation}
where~$d$ is the degree of the graph.
The basis states of $\mathcal{H}_\text{vc}$ are given by
\begin{equation}
        \mathcal{B}_\text{vc}=\{|x,c\rangle \equiv |x\rangle_\text{v}|c\rangle_\text{c} ;
                 x\in\mathbb{Z}_N,c\in\mathbb{Z}_{d}\}
\label{eq:basisVC}
\end{equation}
with cardinality $Nd$.
For a basis state~$|x,c\rangle$,
the index~$x$ identifies the vertex number and $c$ the $c^\text{th}$
state of the coin.
For an edge $e_{xx'}$ we associate the coin state $c$ with the 
edge at $v_x$, and the coin state ${c'}$ with the other end of
the edge at $v_{x'}$.  The values of $c$ and ${c'}$ are
arbitrary but fixed throughout the quantum walk,
to ensure the quantum walker traverses the the graph in a consistent manner.
We define the mapping
\begin{equation}
        \zeta:\mathbb{Z}_N\times\mathbb{Z}_{d}
                \rightarrow\mathbb{Z}_N\times\mathbb{Z}_{d}:
                (x,c)\mapsto\zeta(x,c)=(x',c'),
\label{eq:mapping}
\end{equation}
where $(x,c)$ and $(x',c')$ label each end of $e_{xx'}$.

The unitary quantum walk evolves
by repetition of two steps: a coin toss and a conditional swap.
The coin operator 
\begin{equation}
        C:\mathcal{H}_\text{vc}\rightarrow\mathcal{H}_\text{vc}:|x,c\rangle\mapsto
                \sum_{\tilde{c}\in\mathbb{Z}_{d}}C_{c\tilde{c}}^x|x,\tilde{c}\rangle_\text{c}
\label{eq:Ctransform}
\end{equation}
is a block diagonal matrix with each block labelled by $x$.
The $x$-dependence of the coin matrix allows sufficient freedom in
the quantum walk dynamics for
the quantum coherence properties of the coin to vary between vertices,
for vertices to act as origins and endpoints,
and for vertices to have different degrees from each other.
If $v_x$ has degree $d_x<d$, we require $C_{c\tilde{c}}^x=0$ for all
$\tilde{c}$ values not used to label an edge at $v_x$.
This restricts the coin operator so it only produces states that
have a valid mapping under $\zeta$.
If 
\begin{equation}
        C_{c\tilde{c}}^x=C_{c\tilde{c}}^{x'}\forall x,x',
\end{equation}
we have the special case of a fixed degree graph where
the coin operator is identical for all vertices, as in the walk on a line
or cycle described in the previous sections.

The unitary conditional swap operator is given by
\begin{equation}
        \mathbf{S}:\mathcal{H}_\text{vc}\rightarrow\mathcal{H}_\text{vc}:
                |x,c\rangle\mapsto|x',c'\rangle~,
\label{eq:qswap}
\end{equation}
which updates the position of the walker and the coin state
according to the mapping $\zeta$ in equation~(\ref{eq:mapping}), i.~e., moves 
the walker and coin to the vertex $v_{x'}$ along edge $e_{xx'}$.
We note that, by our stipulation that $c$ and ${c'}$ label
opposite ends of $e_{xx'}$, it follows that $\mathbf{S}=\mathbf{S}^{-1}$, and is thus unitary
as required for quantum evolution.
The sequence of a coin flip and a conditional swap is a transition
over the unit time step, which we denote by unitary $\mathbf{U}=\mathbf{SC}$.
The quantum walk can then be written
\begin{equation}
\ket{\psi(t)}=\mathbf{U}^t\ket{\psi(0)},
\label{eq:qwdgen}
\end{equation}
where $\ket{\psi(t)}$ and $\ket{\psi(0)}$ can be expressed in basis states
as per equation (\ref{eq:basis}).

\subsection{Coined quantum walks on regular lattices}
\label{ssec:lattice}

Analytical solutions of equation (\ref{eq:qwdgen}) can generally only
be obtained for special graphs of fixed degree and high symmetry
\citep[see, for example, ][]{grimmett03a}.
The scattering theory methods of \cite{feldman03a} are worth noting
as a possible exception, though few applications have been presented.
Numerical studies are less constrained, due to the simplicity of
the step by step evolution of quantum walks.
Examples of quantum walks on various graphs of degree
larger than two can be found in \cite{mackay01a,tregenna03a,carneiro05a},
and in the quantum walk search algorithm of \cite{shenvi02a},
see \S\ref{ssec:qwsearch}.
Many of the basic results for quantum walks on the line and cycle,
such as faster spreading, carry over to higher dimensional graphs.
This was shown numerically for regular lattices of degree three and four by
\cite{mackay01a}.  However, higher dimensional coins have a much
wider set of possible types of coin operators, and further numerical
study \citep{tregenna03a}
shows that the quantum speed up is not automatic for all possible
choices.

The role of the lattice symmetry combined with the symmetry
in the dynamics of the quantum walk in determining the spreading
has been further studied and clarified by \cite{krovi05a,krovi06a}.
Most of the examples studied to date have a high degree
of symmetry, and it makes sense to choose a quantum coin operator
that reflects the symmetry in the problem.  
Two examples are worth noting.
First, the DFT (discrete Fourier transform) coin operator is unbiased,
but asymmetric in that you cannot interchange the
labels on the directions without changing the coin operator.
First used by \cite{mackay01a}, for $d=3$ it looks like
\begin{equation}
\mathbf{C}_3^{(D)}=\frac{1}{\sqrt{3}}\left( \begin{array}{ccc}
        1 & 1 & 1\\
        1 & e^{i\omega} & e^{-i\omega}\\
        1 & e^{-i\omega} & e^{i\omega}\\
        \end{array} \right),
\label{eq:DFT3}
\end{equation}
where $e^{i\omega}$ and $e^{-i\omega}$ are the complex cube roots of unity,
For $d=2$, the DFT coin reduces to the Hadamard coin, equation~(\ref{eq:had}),
though this is not the only way to generalise the Hadamard coin to
higher dimensions \citep[see, for example, ][]{tregenna03a,tadej05a}.
Quantum walks using the DFT coin operator have interesting non-classical
properties that have been studied by
\cite{mackay01a,tregenna03a,carneiro05a,krovi05a},
but none have yet provided any quantum algorithms.

The second commonly used coin operator is a highly symmetric coin
based on Grover's diffusion operator
\citep{grover96a}
with elements $2/d-\delta_{ij}$, shown in matrix form for $d=3$,
\begin{equation}
\mathbf{C}_3^{(G)}=\frac{1}{3}\left( \begin{array}{rrr}
        -1 & 2 & 2\\
         2 &-1 & 2\\
         2 & 2 &-1\\
        \end{array} \right).
\label{eq:grov3}
\end{equation}
The Grover coin is biased but symmetric, it is the symmetric
unitary operator furthest from the identity.
It was first used in quantum walks by \cite{watrous98a}, and 
is the key ingredient in the quantum walk searching algorithm
of \cite{shenvi02a}, see \S\ref{ssec:qwsearch}.  \cite{inui03a} 
studied the localisation properties related to searching on a
two dimensional lattice.
\cite{szegedy04a,szegedy04b} introduces a generalisation of the
Grover coin that quantises an arbitrary Markov chain: essentially
this allows for edge weights on the graph, and works for graphs
of variable degree as well as regular graphs.

\subsection{Coined quantum walk on the hypercube: hitting times}
\label{ssec:qwhype}

In one of the few analytical studies of coined quantum walks
in higher dimensions, \cite{moore01a} use a Grover coin for the hypercube,
which has $2^n$ vertices each of dimension $d=n$.
The n-dimensional hypercube is the Cayley graph of
$\mathbb{Z}_2^n$, so the solution follows the same general method as 
for the line and the $N$-cycle, using Fourier transformation of the 
position space.
\cite{moore01a} determine from this solution that the discrete-time
quantum walk on the hypercube of size $n$
has approximate instantaneous mixing times,
\begin{equation}
M_{\text{inst}}(\epsilon) = \left\{t\enspace :\enspace ||P(x,t) - P_u||_{tv} < \epsilon\right\}
\label{minstdef}
\end{equation}
for $t = nk\pi/4$ for all odd $k>0$ and $\epsilon = O(n^{-7/6})$.
This is an improvement over a classical random walk which mixes in
time $O(n\log n)$, but requires measurement of the quantum walk at
exactly the right time (else the walk ``unmixes'' again as it proceeds).
They also consider $\overline{P(x,t)}$, and find using the methods of
\cite{aharonov00a}, that $\overline{M}(\epsilon)$ is exponentially large,
$O(2^n)$.

These results for mixing times are discouraging, but there is another
useful property one can test on a hypercube: hitting times.
\cite{kempe02a,kempe02c} proved that a quantum walk can travel from
one corner of a hypercube to the opposite corner exponentially
faster than a classical random walk.
Kempe defines two different hitting times: instantaneous, where one
measures the destination corner at the optimal time, and concurrent,
where one checks the destination after each step to see if the walker
arrived yet.
This gives us our first taste of non-unitary evolution in a quantum walk,
since measuring the destination node at each step destroys some of the
coherences in the quantum state.
Recent work by \cite{krovi05a} expands on these ideas,
we will describe them in more detail in \S\ref{ssec:hypdec}.
There are other classical algorithms that can cross a hypercube efficiently,
so it does not provide a quantum algorithm with a true advantage
over classical.

\section{Continuous-time quantum walks}
\label{sec:qrwcont}

Continuous-time quantum walks on a discrete lattice have their origins
back as far as \cite{feynman64}.  Their use for quantum algorithms
was first suggested by \cite{farhi98a}, who showed numerically
they can reach the ends of certain network configurations more
efficiently than classical random walks.  A proven exponential
speed up in a quantum algorithm using a continuous-time quantum
walk came a few years later from \cite{childs02a}, which we will
briefly describe in \S\ref{ssec:qwtrees}.
In this section we will first describe the continuous-time quantum
walk dynamics on a general graph, then compare it (numerically) with
the discrete-time walk on the line.
We then give two further examples, on cycles and hypercubes,
that we will need later.

\subsection{Continuous-time quantum walks on general graphs}
\label{ssec:qcongen}

The continuous-time quantum walk naturally works on any undirected graph.
\cite{farhi98a} simply use the adjacency matrix
$\mathbf{A}$, which is symmetric for an undirected graph and therefore
Hermitian, to form the Hamiltonian for the evolution of the quantum state,
\begin{equation}
i\frac{d}{dt}\braket{x}{\psi(t)} =
	\sum_y\bra{x}\mathbf{H}\ket{y}\braket{y}{\psi(t)}.
\label{eq:qcon}
\end{equation}
Here $\mathbf{H}=\gamma\mathbf{A}$, where $\gamma$ is the hopping rate
per edge per unit time, and where $\ket{\psi(t)}$ is now a
vector in the Hilbert space of position (vertices in the graph) only (no coin).
Continuous-time quantum walks achieve what was not possible without a
coin in a discrete-time quantum walk: the traversing of all possible paths
in superposition.  The formal solution of equation~(\ref{eq:qcon}) is
\begin{equation}
\ket{\psi(t)} = 
e^{-i\gamma\mathbf{A}t}\ket{\psi(0)}.
\label{eq:qconsol}
\end{equation}
On the same graph, a continuous-time classical random walk evolves as
\begin{equation}
 \frac{d}{dt}P(x,t) =\gamma\sum_y\left\{ A_{xy}P(y,t) - A_{yx}P(x,t)\right\}.
\label{eq:ccon}
\end{equation}
Comparing equation~(\ref{eq:qcon}) with equation~(\ref{eq:ccon}) shows that in
the quantum walk, the second term, necessary for conservation of
probability, is missing.
We only need a Hermitian operator for quantum evolution, and since 
$A_{xy}=A_{yx}$ guarantees this, we are free to examine this
dynamics as well as that obtained in more direct analogy to
equation~(\ref{eq:ccon}) by using the Laplacian,
\begin{equation}
i\frac{d}{dt}\braket{x}{\psi(t)} =
	\gamma\sum_y\left\{\bra{x}\mathbf{A}\ket{y}\braket{y}{\psi(t)} 
		    - \bra{y}\mathbf{A}\ket{x}\braket{x}{\psi(t)}\right\}.
\label{eq:qcon2}
\end{equation}

For graphs where all the vertices are of the same degree $d$,
the Hamiltonian in equation~(\ref{eq:qcon2}) becomes
$\mathbf{H}=\gamma(\mathbf{A} - d\identity)$,
and the solution to this can be written
\begin{equation}
\ket{\psi(t)}=e^{-i\gamma(\mathbf{A}-d\identity)t}\ket{\psi(0)}.
\end{equation}
Since $\mathbf{A}$ commutes with the identity, the two
terms in the exponential can be factored, giving 
\begin{equation}
\ket{\psi(t)}= e^{-i\gamma\mathbf{A}t}e^{i\gamma d\identity t}\ket{\psi(0)}.
\label{eq:qcphase}
\end{equation}
The factor $e^{i\gamma d\identity t}$ is only an irrelevant global phase
that makes no difference to observable quantities if omitted
\citep{ahmadi02a}, thus equation~(\ref{eq:qcphase}) is equivalent to
equation~(\ref{eq:qconsol}) for graphs of fixed degree.
For graphs of general degree, however, the dynamics with the
second term included will have a different evolution, as \cite{childs03a}
point out.
So far, only graphs of fixed degree have been studied in any detail in
the literature, so the differences between the two versions have not been
explored.

Equation (\ref{eq:qconsol}) looks remarkably similar to the discrete
case, equation~(\ref{eq:qwdgen}): both can be written in the form
\begin{equation}
\ket{\psi(t)}=\mathbf{U}^t\ket{\psi(0)},
\end{equation}
with unitary operator $\mathbf{U}=\mathbf{SC}$ for the discrete-time walk
and $\mathbf{U}=e^{-i\gamma\mathbf{A}}$ for the continuous-time walk.
But, unlike the classical case, where the limit of the discrete-time
walk as the time step goes to zero can be taken in a way that gives the
continuous-time walk, in the quantum case
the similarity is deceptive.  The discrete and continuous walks have
Hilbert spaces of different sizes, since the continuous-time quantum walk
has no coin: the continuous-time walk therefore cannot be the limit of
the discrete-time walk unless it can be restricted to a subspace that
excludes the coin.  \cite{strauch05a,strauch06a} shows how to do this for the
walk on the line: it turns out the appropriate limit of the discrete
time quantum walk produces two copies of the continuous-time quantum
walk, one for each of the two coin degrees of freedom.
Alternatively, if one adds a coin space to 
the continuous-time quantum walk, a quantum dynamics is obtained
that is more obviously the continuum limit of the discrete-time
coined quantum walk, see \cite{szegedy04b,childs04a}.
The quantised Markov chain formalism introduced by \cite{szegedy04a} is yet
another convenient way of constructing quantum walks in a parallel
manner to classical random walks \citep[e.g., see][]{weiss94}.
The stochastic matrix governing the Markov chain corresponds to the
adjacency matrix in the graph-based description above.

\subsection{Continuous-time quantum walk on the line}
\label{ssec:qconline}

\begin{figure}
  \begin{center}
    \resizebox{0.6\textwidth}{!}{\includegraphics{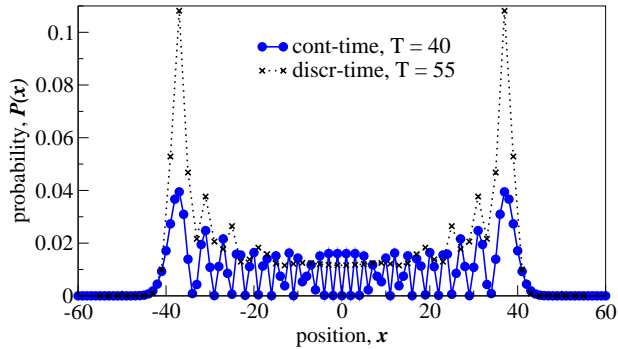}}
  \end{center}
  \caption{Comparison of continuous-time (dots) and discrete-time (crosses)
	quantum walks on a line starting at the origin
	after 40 and 55 steps respectively.
	The discrete-time walk has initial coin state
	$(\ket{-1} +i\ket{1})/\sqrt{2}$, and only
	even points are shown since odd points are unoccupied at even
	time steps (and vice versa).  All points are shown for the
	continuous-time walk, which has hopping rate $\gamma=0.5$.}
  \label{fig:convsdis}
\end{figure}

A comparison of the simple case of the walk on the line 
will show us the main similarities and differences
between discrete-time and continuous-time quantum walks.
Equation (\ref{eq:qcon}) has a straightforward solution in terms of
Bessel functions, 
\begin{equation}
	\ket{\psi(t)} = \sum_{x=-\infty}^{+\infty}(-i)^xJ_x(t)\ket{x},
\end{equation}
where $J_x(t)$ is the Bessel function of order $x$.
Using a hopping rate of $\gamma=0.5$ so the total
hopping probability per unit time is one, after 40 units of time
we obtain the probability distribution shown in figure \ref{fig:convsdis}.
The discrete-time quantum walk evolved for 55 time steps is also
shown for comparison.
We can see that the shapes of the distribution are comparable,
the difference in the height of the peaks is due largely to the
continuous-time walk having support on both odd and even sites,
while the discrete-time walk is
restricted to sites with the same parity as the time step.
Nonetheless, the continuous-time walk has a very oscillatory nature
about it, especially in the central region.  The two types of walks
propagate at somewhat different speeds, as evidenced
by the different time instants at which they have the same width.
Both spread linearly, just with different constant prefactors, so 
both show a quadratic speed up in their spreading compared to 
classical random walks.  For the continuous-time quantum walk starting
at the origin, for asymptotically large times $T$,
\begin{equation}
\sigma^2(T) = \langle x^2\rangle = T^2/\gamma.
\label{eq:sigma2c}
\end{equation}
The continuous-time quantum walk on the line is always symmetric,
if the hopping rate is the same for both directions,
this is the main difference with the discrete-time walk, where the
coin generally skews the walk unless special choices are made for
the initial state.

\subsection{Continuous-time quantum walk on a $N$-cycle and other
circulant graphs}
\label{ssec:qconcyc}

The continuous
time quantum walk on a $N$-cycle is straightforward to analyse.
The adjacency matrix $\mathbf{A}$ of the $N$-vertex cycle graph
is a circulant matrix, it has eigenvalues $\lambda_{x} = 2\cos(2\pi x/N)$
with corresponding eigenvectors $\ket{b_{x}}$, 
where $\braket{y}{b_{x}} = \frac{1}{\sqrt{N}}\exp(-2\pi i xy/N)$,
for $x = 0,1,\ldots,N-1$. 
Taking the initial state of the quantum walk to be $\ket{\psi(0)} = \ket{0}$,
then $\ket{\psi(t)} = e^{-it \mathbf{H}}\ket{0}$ can be solved by 
decomposing $\ket{0}$ in terms of the eigenvectors $\ket{b_{x}}$, giving
\begin{equation}
\ket{\psi(t)}
        = \frac{1}{\sqrt{N}} \sum_{x=0}^{N-1} e^{-it\lambda_{x}}\ket{b_{x}}.
\end{equation}

One can apply the same notions of mixing times for a
continuous-time quantum walk as for the discrete-time walk.
Like the discrete-time quantum walk on a cycle, the 
probability distribution $P(x,t)$ of the continuous-time walk 
does not mix asymptotically, and is known to have exact instantaneous mixing 
only for a few special cases, $N=3$ and $N=4$ \citep{ahmadi02a},
i.e. even less cases than are known for the discrete-time walk
on the cycle as mentioned in \S\ref{ssec:pd}.
We can define the continuous-time version of
equation~(\ref{eq:meanP}), the average probability
\begin{equation}
\overline{P(x,T)} = \frac{1}{T} \int_{0}^{T}dt\, P(x,t)
\end{equation}
where $P(x,t) = |\braket{x}{\psi(t)}|^2$.
While this does mix asymptotically, like the discrete-time walk, in general the
limiting distribution is not uniform: it retains one or more peaks
reflecting the initial state.
Notice that the properties of continuous-time walks on cycles are not
dependent on whether the cycle has an odd or even number of nodes,
unlike the discrete-time quantum walk.

\cite{adamczak03a} show that 
continuous-time walks on cycles are nearly uniform mixing, i.e., if one
relaxes the condition on how well they they approach the uniform distribution
from an arbitrarily small $\epsilon$, they satisfy
\begin{equation}
\overline{M}(1/4N) = \min\left\{T\enspace|\enspace\forall\enspace t>T:
		||\overline{}P(x,t) - 1/N||_{tv} \le \frac{1}{4N}\right\}
		\sim O(N).
\end{equation}
\cite{adamczak03a} extend their analysis of mixing in continuous-time
quantum walks to a wider variety of circulant graphs, this is continued in
\cite{lo05a}: mixing to non-uniform distributions turns out to be
the norm for these types of graphs.  \cite{carlson06a} identify graphs
where, with the use of edge weights, any distribution on the graph
can be obtained (universal mixing).  They also identify another class
of graphs that are instantaneous uniform mixing, the claw or star graphs.

\subsection{Continuous-time quantum walk on the hypercube}
\label{ssec:hypercube}

Our next example of a continuous-time quantum walk is on a
$n$-dimensional hypercube.
\cite{moore01a} analytically solved 
the continuous-time quantum walks on the hypercube as well
as the discrete-time walk (see \S\ref{ssec:higherD}).
The analysis for the continuous-time walk makes use of the hypercube's
product graph structure.
We label the vertices with $n$-bit strings, with edges connecting
those pairs of vertices that differ in exactly one bit.
Then, using the Pauli matrix, $\sigma_x$, which is the bit flip operator,
\begin{equation}
\sigma_x = \left(\begin{matrix} 0 & 1 \\ 1 & 0 \end{matrix} \right),
\end{equation}
the adjacency matrix can be decomposed into the sum
\begin{equation}
\mathbf{A} = \sum_{j=1}^n \identity \otimes \cdots \otimes \sigma_x \otimes 
   \cdots \otimes \identity\enspace,
\label{eq:hypA}
\end{equation}
where the $j$th term in the sum has $\sigma_x$ as the $j$th factor
in the tensor product.
Each term thus flips the bit in the vertex label corresponding to
traversing the edge to the appropriately labelled neighbouring vertex.
Using $\mathbf{H} = \gamma\mathbf{A}$ for the quantum walk,
we have\footnote{\cite{moore01a} use $U = e^{iH}$, this is just the
complex conjugate and makes no difference to observable quantities.}
\begin{eqnarray}
\mathbf{U}^t &=& e^{-i\mathbf{H}t} = \prod_{j=1}^n \identity \otimes \cdots \otimes e^{-i\gamma t\sigma_x} \otimes \cdots
\otimes \identity = \left[e^{-i\gamma t\sigma_x} \right]^{\otimes n} \nonumber\\
& = & \left(\begin{matrix}\cos(\gamma t) & -i~\sin(\gamma t) \\ -i~\sin(\gamma t) & \cos(\gamma t) \end{matrix}\right)^{\otimes n}\enspace.
\end{eqnarray}
Applying $\mathbf{U}^t$ to the initial state
$\ket{\psi(0)} = \ket{0}^{\otimes n}$, we have
\begin{equation}
\mathbf{U}^t \ket{\psi(0)} = \left[ \cos(\gamma t) \ket{0}
  - i~\sin(\gamma t) \ket{1} \right]^{\otimes n}
\end{equation}
which corresponds to a uniform state exactly when $\gamma t$
is an odd multiple of $\frac{\pi}{4}$.
The choice of $\gamma$ is thus pivotal for making a fair comparison
of the instantaneous mixing time.  \cite{moore01a} choose
$\gamma=1/n$, so that the total probability of making a hop to any
neighbouring site per unit time is unity.
With this choice, the continuous-time quantum walk mixes exactly
instantaneously in time $O(n)$, a logarithmic improvement over
$O(n\log n)$ for the mixing time of a classical random walk.
They also showed that for the time-averaged
mixing time given by equation (\ref{eq:meanmixdef}),
the continuous-time walk never mixes.
These results can be compared with the discrete-time walk on
the hypercube, which has approximate instantaneous mixing
(instead of exact), and $\overline{M}(\epsilon)$ mixing
in exponential time.

The continuous-time quantum walk on the hypercube also hits the opposite
corner in linear time, exponentially faster than a classical random walk:
we will obtain this result in \S\ref{ssec:chypdec} as
a special case of the calculation of the effects of decoherence.
As with the coined quantum walk, this exponentially fast hitting time
is highly dependent on the symmetry.
\cite{keating06a} apply results from localisation theory \citep{anderson58a}
to argue that this behaviour is exceptional, and the norm on less regular
graphs is a quantum walk that tends to stay near its starting state.

\section{Algorithms using quantum walks}
\label{sec:qrwalgo}

Starting with Shor's algorithm for
factoring large numbers \citep{shor95a}, many of the quantum
algorithms found so far belong to
the same family, based on the use of Fourier transforms to identify a
hidden subgroup \citep[provides a recent review]{lomont04a}.
This works well (can be exponentially better than known classical methods)
for Abelian groups, but extending
the method to non-Abelian groups, where some of the notorious hard 
problems, such as graph isomorphism, lie, is proving tricky.
An obvious place to look for new
ideas is where classical algorithms are having the most success, to
see if a quantum version could be even faster.  Randomised algorithms
are one such arena, providing the best known algorithms for approximating
the permanent of a matrix \citep{jerrum01a}, finding satisfying
assignments to Boolean expressions ($k$SAT with $k>2$) \citep{schoning99a},
estimating the volume of a convex body \citep{dyer91a},
and graph connectivity \citep{motwani95}.
Classical random walks also underpin many standard methods in computational
physics, such as Monte Carlo simulations.
Expanding the repertoire of methods for quantum algorithms was the
motivation behind the recent upsurge of interest in quantum walks.
The first proper algorithms using quantum walks 
appeared from \cite{childs02a} and \cite{shenvi02a}, and more have since
followed: for a short survey, see \cite{ambainis04a}.
We will briefly describe the first two quantum walk algorithms, since they
are of two distinct generic types, and later algorithms are mostly
variants of the quantum walk search outlined in the next section.

\subsection{Quantum walk searching}
\label{ssec:qwsearch}

Quadratically faster spreading, as described in \S\ref{ssec:2dline},
isn't a quantum algorithm,
but quantum searching of an unsorted database (e.g., finding the name 
corresponding to a given number by searching a telephone directory)
is the reverse process: start in a uniform superposition over
the whole database and home in on the marked item.
The first quantum algorithm for this
problem is due to \cite{grover96a}, using a method of amplitude 
amplification, from which a quadratic speed up over classical
searching is obtained.  It can be shown that a quadratic speed up is
the best possible improvement for this problem \citep{bennett97a}.
A classical search of an unsorted database
potentially has to check all $N$ entries in the database, and on average
has to check at least half.  A quantum search only needs to make
$O(\sqrt{N})$ queries, though the queries ask for many answers in superposition.
\cite{shenvi02a} showed that a quantum walk can also search an unsorted database
with a quadratic speed up.  
They represent the database by the vertices of a graph with a regular structure
(lattice, hypercube...) and start with the quantum walker in a superposition
of all positions on the vertices.  The quantum walk proceeds
using a Grover coin operator of appropriate dimension at every
vertex except the marked item.  For the marked item,
almost any other coin operator can be used, this is enough to 
break the symmetry of the quantum walk and cause the walker to converge
on the marked vertex in $\sim \pi/2\sqrt{N/2}$ steps.

Since then, several variations on quantum 
searching for a single item have been analysed,
all essentially searching for a set of marked items of some sort
\citep{magniez05a,childs03b,ambainis03a}.
Spatial searching, where it is also counted as a cost to move from one
item in the database to the next, is also faster with a quantum
algorithm, and curiously, the continuous-time walk finds this a
little harder than the discrete-time walk: in low dimensions it needs
a coin too \citep{childs03a,childs04a}.

\subsection{``Glued trees'' algorithm}
\label{ssec:qwtrees}

\begin{figure}
    \begin{center}
    \resizebox{0.5\textwidth}{!}{\includegraphics{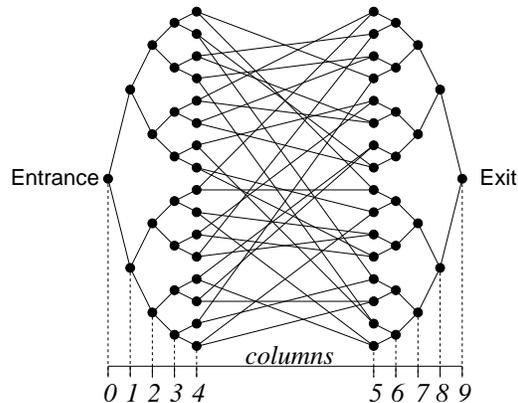}}
    \end{center}
            \caption{``Glued trees'' graph used in the algorithm of
                \cite{childs02a}.  Example shown is for $N=4$, with
                $2N+2=10$ columns labelled at the bottom of the figure,
                and $2(2^{(N+1)}-1)=62$ nodes.
                The task is to travel from entrance
                to exit without getting lost in the randomly joined middle
                section of the graph.  The gap between columns 4 and 5 is for
                clarity in the figure and is not significant in the algorithm.}
    \label{fig:tree4}
\end{figure}

\cite{childs02a} proved that a continuous-time quantum walk
could perform exponentially faster than any classical algorithm when 
finding a route across a particular sort of network,
shown in figure~\ref{fig:tree4}.  This is a rather artificial problem,
but proves in principle that quantum walks are a powerful tool.
The task is to find your way from the entrance node to the exit node,
treating the rest of the network like a maze where you cannot see the other
nodes from where you stand, only the choice of paths.  It is easy to tell
which way is forward until you reach the random joins in the centre.
After this, any classical attempt to pick out the correct path to the exit 
get lost in the central region and takes exponentially long, on average,
to find the way out.  A quantum walk, on the other hand,
travels through all the paths in superposition, and the quantum interference
between different paths
allows the quantum walker to figure out which way is forward right up to
the exit, which it finds in time proportional to the width of the network.
The proof is quite technical, involving oracles and consistent colourings
of the network, and the ability to simulate a continuous-time quantum
walk on a discrete quantum computer.  We refer the reader to the original
paper for the details \citep{childs02a}, since we do not need them
here for our discussions of decoherence.

Discrete and continuous-time quantum walks are generally expected to 
have the same computational power.  They give broadly similar results for 
algorithmic properties such as mixing times and hitting times
in those cases where both forms have been applied to the same problem,
with the possible exception of spatial search \citep{childs03a},
which, as noted above, requires a coin for maximum efficiency
even in the continuous-time walk in low dimensions \citep{childs04a}.
For a discussion of how both discrete and continuous-time quantum walk
implementations would use roughly the same computational resources
for the ``glued trees'' problem, see \cite{kendon03c}.

\section{Decoherence in coined quantum walks}
\label{sec:qrwdiscdec}

As we have seen, while there are many similarities between discrete
and continuous-time quantum walks, there are also some distinctly
different behaviours, and this turns out to be true also 
under the influence of decoherence.  We will thus devote a
separate section to each type, starting here with coined quantum walks
and following in \S\ref{sec:qrwcontdec} with continuous-time quantum walks.
We are going to take a very broad view of decoherence in this review as
any dynamics that tends to remove the quantum coherences in some way,
be it unwanted (as in environmental decoherence in an experimental system),
intentional (to tune the properties of the quantum walk), or a
byproduct of some other operation, such as measurement.
One of the earliest uses of non-unitary quantum walks is
measurement-based, by \cite{aharonov92a}.

Quantum walks are a very broad class of quantum dynamics, with overlap
into related areas such as quantum graphs \citep{kottos97a}.  One way to 
justify a particular quantum dynamics as being a ``quantum walk'' is
to see if it turns into a classical random walk when decohered.
Since classical random walks are also a very broad class of dynamics,
this gets reasonably close to a workable definition of a quantum walk,
at least for the discrete-time case.
Many early studies of quantum walks took the trouble
to show numerically that, for specific cases, their quantum walks
decohered into classical random walks, \cite[e.g., see][]{mackay01a}
who dephase the coin in their study of a quantum walk on the line.
A more systematic treatment of the quantum to classical transition
in a general quantum walk appears in \cite{kendon04a}, emphasising the
importance of demonstrating that quantum walks exhibit
both wave (pure quantum) and particle (decohered) dynamics
and, for a non-unitary quantum walk, being able to interpolate between 
these two modes of behaviour by turning the decoherence up or down.
\cite{kosik06a} explicitly calculate the cases of Grover and DFT coins
on a Cartesian lattice ($\mathbb{Z}^d$), with random phase shifts
applied to the coin to induce decoherence, rather than measurements.  
They show that the resulting distributions match the expected classical
random walks (which may be biased if the initial state or coin operator 
is biased, compare \S \ref{ssec:2dline}).

Decoherence is usually modelled as a non-unitary evolution of the
quantum walk, so we will need some extra
formalism for mixed states using density matrix operators.
Following the notation in \S\ref{ssec:higherD},
the (time-dependent) density operator
\begin{equation}
        \bm\rho = \sum_{x,c}\sum_{x',c'}\rho_{xc,x'c'}|x,c\rangle\langle x',c'|,
\end{equation}
is a positive ($\bm\rho=\bm\rho^\dagger$ with positive real spectrum),
unit-trace, bounded linear operator on $\mathcal{H}_\text{vc}$,
in the basis $\mathcal{B}_\text{vc}$, equation~(\ref{eq:basisVC}).
In the decomposition of $\bm\rho$ into the computational basis,
the rows and columns of $\bm\rho$ are indexed by $xc$ and $x'c'$, which run
over the position states and coin states of the Hilbert space.
The state is pure iff $\bm\rho^2=\bm\rho$.
A typical initial condition is $\bm\rho(t=0)=|0,0\rangle\langle 0,0|$
corresponding to the walker starting at vertex $v_0$ carrying a coin in the 
state labelled zero.
In general the density operator is mapped to a new density operator via
a completely positive (CP) map
\begin{equation}
\mathcal{U}:\bm\rho\mapsto\mathcal{U}\bm\rho
\end{equation}
The CP map $\mathcal U$ performs 
both the coin flip and the conditional swap over one time step.
More explicitly, we can write
\begin{equation}
        \mathcal{U}\bm\rho=\sum_{i\in\Theta} \mathbb{U}_j^\dagger \bm\rho \mathbb{U}_j,
        \text{\null\hspace{1em}\null}
        \sum_{i\in\Theta} \mathbb{U}_j^\dagger \mathbb{U}_j = \identity,
\label{eq:Udecomposition}
\end{equation}
with $j$ an index of non-unitary evolutionary `instances'
and $\mathbb{U}_j$ the corresponding
Kraus operator.
These instances may be discerned by a measurement record, with $j$ the
record index.  The cardinality of $\Theta$ can be finite, countably
infinite, or even uncountable.  In the case of unitary evolution,
$\Theta$ has a cardinality of one, so there is a single,
unitary $\bm U$ for which
$\bm\rho\mapsto\mathcal{U}\bm\rho = \bm U \bm\rho \bm U^\dagger$.
Unitary quantum walk evolution can be expressed as
\begin{equation}
        \bm\rho(t) = \mathcal{U}^t \bm\rho(0),\; \mathcal{U}\equiv\mathcal{SC},
                \; \mathcal{SC}\bm\rho\equiv \bm{SC}\bm\rho \bm C^\dagger \bm S^\dagger~,
\label{eq:Uunitary}
\end{equation}
where, for the discrete-time walk, we assume $t\in\mathbb{Z}$.
Thus, for the unitary walk, a single step is given by $\bm U=\bm{SC}$,
while for the non-unitary walk we can either add an extra non-unitary
operation (measurement) $\{\mathbb{P}_j\}$,
\begin{equation}
        \mathcal{U}:\bm\rho\mapsto\sum_{i\in\Theta}\mathbb{P}_j\bm{SC}\bm\rho \bm C^\dagger \bm S^\dagger \mathbb{P}_j^\dagger,
\label{eq:Unonunitary}
\end{equation}
or, replace the coin and/or the shift operators by non-unitary operators,
depending on what sort of decoherence or measurements we are considering.

Most of this section consists of unpacking equation (\ref{eq:Unonunitary})
into specific cases, and analysing the effects.  One may ask, for example,
how quickly the quantum walk becomes classical as the decoherence is 
increased.  Are quantum walks sensitive to small amounts of decoherence,
or are the quantum effects robust under environmental disturbance?
The first studies of decoherence, beyond merely verifying the classical
limit is a classical random walk, were analysis of the likely errors
in proposed experimental implementations in
\cite{travaglione01a,sanders02a,dur02a}.  These, too, are focused on the
properties of the pure quantum walk, we provide a short summary in
\S\ref{sec:qrwphys}.  The first consideration of the algorithmic
properties of partly decohered quantum walks were by \cite{kendon02c}, who
used numerical simulation on a variety of discrete coined quantum walks.
We will briefly review their results for the walk on the line, to
provide an overview of the typical effects of decoherence, then
examine some of the analytical calculations that confirm and
expand these initial observations.
An alternative approach is to model the entire system of quantum walker
plus environment, we include examples of this in \S\ref{ssec:manycoin},
where the different  methods of decohering the coin are compared.

\subsection{Effects in the walk on the line}
\label{ssec:lindecnum}

First we unpack the superoperator notation of equation~(\ref{eq:Unonunitary})
into the specific case of decoherence events or measurements
occurring independently at each time step,
\begin{equation}
\bm\rho(t+1) = (1-p)\mathbf{SC}\bm\rho(t)\mathbf{C}^{\dag}\mathbf{S}^{\dag}
          + p\sum_j\mathbb{P}_j\mathbf{SC}\bm\rho(t)
		\mathbf{C}^{\dag}\mathbf{S}^{\dag}\mathbb{P}_j^{\dag}.
\label{eq:decdyn}
\end{equation}
Here $\mathbb{P}_j$ is a projection that represents the action of the
decoherence and $p$ is the probability of a decoherence event
happening per time step, or, completely equivalent mathematically,
to a weak coupling between the quantum walk system and some
Markovian environment with coupling strength $p$.
Equation (\ref{eq:decdyn}) lends itself readily to numerical simulation since
$\bm\rho$, $\mathbf{S}$ and $\mathbf{C}$ can be manipulated as complex
matrices, while the $\mathbb{P}_j$ generally remove some or all
of the off-diagonal entries in $\bm\rho$.
\cite{kendon02c} took equation (\ref{eq:decdyn})
and evolved it numerically for various choices of $\mathbb{P}_j$,
projection onto the position space, projection into the coin space
in the preferred basis $\ket{\pm 1}$, and projection of both
coin and position.
Motivated by the likely form of experimental errors,
they also modelled an imperfect Hadamard 
by applying a Gaussian spread of standard deviation $\sqrt{p}\pi/4$
about the perfect value of $\pi/2$ implicit equation~(\ref{eq:had}),
compare \cite{mackay01a}.
\cite{shapira03a} later modelled imperfect quantum walk operations
such as this in more detail, still numerically, and drew the same
conclusions, while \cite{konno04a} treated the general case of a
randomised coin, showing analytically that the classical random
walk is obtained.
An imperfect shift by the walker has been studied in \cite{dur02a},
see \S\ref{sec:qrwphys}, and also (in the form of
broken links) by \cite{romanelli04a}, see \S\ref{ssec:lindec}.

\begin{figure}
            \begin{center}
            \resizebox{0.6\columnwidth}{!}{\includegraphics{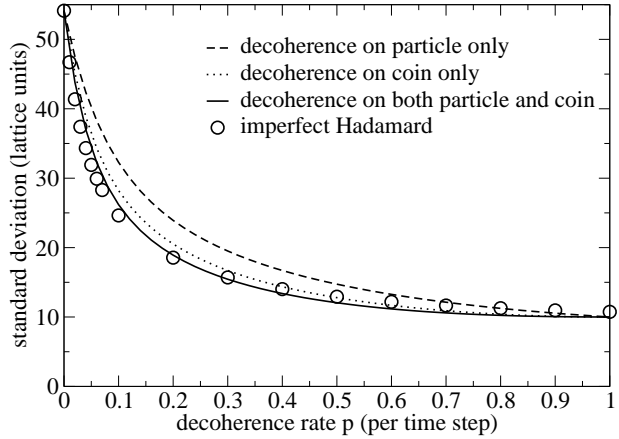}}
            \end{center}
            \caption{Standard deviation $\sigma(T,p)$ of the position on
                a line for different models of decoherence, for $T= 100$
                time steps, \citep[from][]{kendon02c}.}
            \label{fig:dec100-all}
\end{figure}

To quantify the change in behaviour in the walk on the line, we can 
calculate the standard deviation, equation (\ref{eq:sigma2}), which is
now also a function of the decoherence rate.
In each of these cases \cite{kendon02c}
found the same general form for the decay of 
$\sigma(T,p)$ from the quantum to the classical value,
with small differences in the rates, as shown in figure~\ref{fig:dec100-all}.
The slope of $\sigma(T,p)$ is finite as $p\rightarrow 0$ and zero
at $p = 1$.
The decay of the spreading rate (as quantified by $\sigma(T,p)$),
is much as one might have expected given quantum states are generally
fragile in the face of environmental disturbance.
The interesting feature is seen in the shape of the
distribution of the position as decoherence begins to
take effect.  
\begin{figure}
  \begin{center}
    \resizebox{0.7\textwidth}{!}{\includegraphics{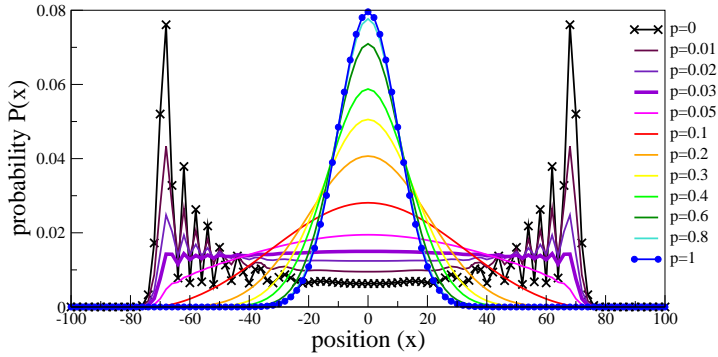}}
  \end{center}
  \caption{A quantum walk on a line of 100 steps is progressively
        decohered by random measurements with probability $p$ per
        time step as given in the key.  for $p=0.03$, an approximately
        `top hat' distribution is obtained.}
  \label{fig:qwdec}
\end{figure}
The changing shape as decoherence is increased is shown in figure~\ref{fig:qwdec}.
Notice the good approximation to a top-hat distribution between
$\pm T/\sqrt{2}$ that appears for
$p=0.03$.  For computational physicists who use random walks to sample
distributions, this is a very desirable feature, since it provides uniform
sampling over a specific range.
Moreover, this result is only obtained when decohering the position,
compare the three examples in figure \ref{fig:200pdf}.
\begin{figure}
            \begin{center}
            \resizebox{0.6\columnwidth}{!}{\includegraphics{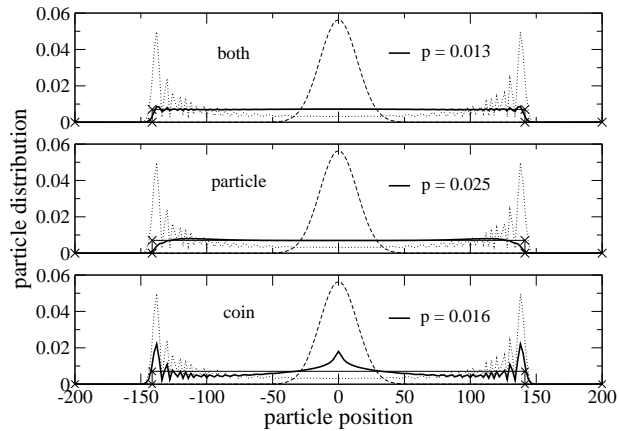}}
            \end{center}
            \caption{Distribution of the position 
                for a quantum walk on a line after $T=200$ time steps.
                Pure quantum (dotted), fully classical (dashed),
                and decoherence at rate shown
                on part of system indicated by key (solid).
                Uniform distribution between
                $-T/\sqrt{2} \le x \le T/\sqrt{2}$ (crosses) also shown,
		\citep[from][]{kendon02c}.}
            \label{fig:200pdf}
\end{figure}
The optimal decoherence rate $p_u$ can be obtained
by calculating the total variational distance,
equation (\ref{eq:tvd}), between the actual distribution
and an ideal top hat distribution,
The optimum decoherence rate depends on the number of steps in the walk,
$p_uT \simeq 2.6$ for decoherence on both coin and position
and $p_uT \simeq 5$ for decoherence on the position only (obtained
numerically in \cite{kendon02c}).  Intriguingly, recent work by
\cite{maloyer06a} shows that the optimal top-hat occurs for
the decoherence rate that just removes all the quantum correlations
by the end of the quantum walk.

Decoherence on the coin only does not produce a top hat distribution,
instead the distribution develops a cusp as it passes from quantum to
classical.  The work of \cite{lopez03a} provides some insight into
why decoherence on the coin and on the position produce qualitatively
different results.
\citeauthor{lopez03a} present an elegant analysis of decoherence in 
the quantum walk using a discrete Wigner function
to bring out the pattern of correlations and their decay when
the coin is subject to decoherence.
Their decoherence model follows much the same approach
as \cite{brun02c}, which will be described in \S\ref{ssec:lincoindec}.
In their study, \cite{lopez03a} restrict the dynamics to cycles.
However, most of their results are for a short numbers of time steps
such that the walk has not joined up round the cycle, and the results
are identical to the walk on the line.
Wigner functions represent quantum dynamics in a
quantum phase space, thus showing the behaviour of both position
and momentum degrees of freedom in the same picture.  This brings out
beautifully that while decohering the coin state is sufficient to
reduce a quantum walk on a cycle to a classical walk,
if the walker started in a superposition of two position states, this
remains untouched by the decoherence.  Such a walker with a decohered
coin performs a superposition of two classical random walks, each starting
at a different position.  The converse does not apply, because the position
is conditioned on the state of the coin, equation (\ref{eq:shift}),
thus decohering the position decohers the coin with it.

Our first example of decoherence has given us a good guide for what
to expect in general: besides the rapid degradation
of quantum behaviour, there are interesting effects in particular
ranges of low decoherence.   These effects are not algorithmically significant,
that is, they don't alter the scaling of quantum processing,
but nevertheless may be useful in practice, and
represent intriguing and complex behaviour worthy of study in 
their own right.
For practical purposes (assuming that one day we have a quantum
computer available), the improved top hat profile might still be a useful
optimisation to get the most out of the computational resources.

\subsection{Dephasing the coin in the walk on the line}
\label{ssec:lincoindec}

In the next two sections we will consider the analytical treatment of
decoherence in the walk on the line in some detail.
First we consider decoherence only on the coin.
This will also provide an example of how to use Fourier
transforms to simplify the dynamics and obtain the solution.
We follow the method of \cite{brun02c,brun02a} to
solve equation~(\ref{eq:Unonunitary}) for the
coined quantum walk on a line with dephasing applied to the coin.
We thus have $\mathbf{S}$ given by equation~(\ref{eq:Sdef}),
\[
\mathbf{S}\ket{x,c} = \ket{x+c,c},
\]
where $\ket{x}$ and $\ket{c}$ are basis states,
and $\mathbf{C}$ by equation~(\ref{eq:had}) 
\[
\mathbf{C}_2^{(\text{H})}=\frac{1}{\sqrt{2}}\left( \begin{array}{rr}
        1 & 1 \\
        1 & -1 \\
        \end{array} \right).
\]
Transforming from the position basis $\ket{x}$
to the Fourier basis $\ket{k}$ such that,
\begin{equation}
\ket{k,c} = \sum_x e^{ikx}\ket{x,c}
\label{eq:ftline}
\end{equation}
we find that
\begin{equation}
\mathbf{U}_k\ket{k,c} = \mathbf{C}_k\ket{k,c}
\end{equation}
where $\mathbf{C}_k$ acts only on the coin degrees of freedom,
\begin{equation}
\mathbf{C}_k=\frac{1}{\sqrt{2}}\left( \begin{array}{rr}
        e^{ik} & e^{ik} \\
        e^{-ik} & -e^{-ik} \\
        \end{array} \right).
\end{equation}
We now move to density matrix formulation, 
for a walk starting at the origin in coin state $\ket{\psi_0}$
(not necessarily a basis state),
\begin{equation}
\bm\rho_0 = \ket{0,\psi_0}\bra{0,\psi_0} =
	\int\frac{dk}{2\pi} \int\frac{dk'}{2\pi}
		\ket{k}\bra{k'}\otimes\ket{\psi_0}\bra{\psi_0},
\end{equation}
where $\ket{k,\psi_0}$ is the tensor product $\ket{k}\otimes\ket{\psi_0}$.
Using equation~(\ref{eq:Unonunitary}) to introduce a decoherence
operator, we can thus write
\begin{equation}
\bm\rho(t) = \int\frac{dk}{2\pi} \int\frac{dk'}{2\pi}
	\ket{k}\bra{k'} \otimes \mathcal{P}_{kk'}^t\ket{\psi_0}\bra{\psi_0},
\label{eq:depcoin}
\end{equation}
where 
\begin{equation}
\mathcal{P}_{kk'}\ket{\psi_0}\bra{\psi_0} =
	\sum_j \mathbb{P}_j\mathbf{C}_k\ket{\psi_0}
		\bra{\psi_0}\mathbf{C}_{k'}^\dagger\mathbb{P}_j^\dagger,
\label{eq:coinoise}
\end{equation}
and the set of projectors $\{\mathbb{P}_j\}$ is the decoherence acting
on the coin that we will need to specify in order to carry out
explicit evaluation of equation~(\ref{eq:depcoin}).

While it will in general be difficult to evaluate $\bm\rho(t)$ explicitly,
the second moment will give us
the main information we need to determine how fast the walk spreads.
Calculating the moments allows considerable simplification even
before specifying the form of decoherence,
\begin{equation}
\langle x^m\rangle = \sum_x x^mP(x,t) = \sum_x x^m
	 \int\frac{dk}{2\pi} \int\frac{dk'}{2\pi}
	\braket{x}{k}\braket{k'}{x}
	Tr_c\left[\mathcal{P}_{kk'}^t\ket{\psi_0}\bra{\psi_0}\right],
\end{equation}
where $Tr_c[.]$ traces over the coin degrees of freedom to aggregate
the probability over the different coin states.
From equation~(\ref{eq:ftline}), $\braket{x}{k} = e^{ikx}$, and we may
carry out the sum over $x$, giving
\begin{equation}
\langle x^m\rangle = \frac{(-1)^m}{2\pi}
	 \int dk \int dk' \delta^{(m)}(k-k')
	Tr_c\left[\mathcal{P}_{kk'}^t\ket{\psi_0}\bra{\psi_0}\right],
\end{equation}
where $\delta^{(m)}(k-k')$ is the $m$th derivative of the delta function.
This can then be integrated by parts.

\cite{brun02a} chose pure dephasing for the form of the decoherence on the coin,
the coin projectors in equation (\ref{eq:coinoise}) are thus
\begin{equation}
\mathbb{P}_{\pm 1}=\frac{1}{\sqrt{2}}\left( \begin{array}{rr}
        e^{\pm i\theta} & 0 \\
        0 & e^{\mp i\theta} \\
        \end{array} \right).
\end{equation}
This can be solved explicitly, giving in the large $t$ limit,
\begin{equation}
\langle x^2\rangle - \langle x\rangle^2 \simeq
	t(\cot^22\theta + \csc^22\theta) + O(1),
\end{equation}
where $\theta$ measures the strength of the dephasing, $\theta=\pi/4$
being complete dephasing (classical walk) and $\theta=0$ is a pure
quantum walk (the approximation above is not valid for exactly $\theta=0$).
This shows that even a small amount of decoherence renders the
quantum walk classical in the sense that the standard deviation of the
position scales as $\sqrt{t}$, but with a larger prefactor the smaller
the dephasing, $\cot^22\theta \sim 1/\theta^2$ for small $\theta$.

We will return to decohering the coin in \S\ref{ssec:manycoin}
where a variety of other ways to decoher the coin will
be considered and compared.  Next, we will complete the picture for
decohering a walk on the line with a calculation of the
effects of decoherence in the position.

\subsection{Decohering the position in the walk on the line}
\label{ssec:lindec}

When decoherence acts on the position degrees of freedom,
the simplification offered by the factorisation of the dynamics in
Fourier space no longer helps: we would have to apply the
Fourier transform to the decoherence superoperator too.
The following calculations of the asymptotic
behaviour for small $p\ll 1$ illustrate how real space (path counting)
methods may be used instead.
Following \cite{kendon02b}, we calculate $\sigma(T,p)$ analytically
for $pT \ll 1$ and $T \gg 1$ for the case where the $\{\mathbb{P}_j\}$ are the
projectors onto the preferred basis $\{\ket{a,x}\}$, i.e., decoherence
affecting both walker and coin simultaneously.  Since, as already explained,
decohering the position also decohers the coin, it makes only a small
difference whether or not we also decoher the coin explicitly.

The probability distribution for finding the
walker in the state \ket{a,x} in the presence of decoherence can be written,
\begin{equation}
P(x,a,T,p) = (1-p)^TP(x,a,T) + p(1-p)^{T-1}P^{(1)}(x,a,T) + \dots ,
\label{eq:PxapTgen}
\end{equation}
where $P(x,a,T)$ is the distribution obtained for a perfect walk and
$P^{(i)}(x,a,T)$ is the sum of all the ways to have exactly $i$ noise
events For example, we can write
\begin{equation}
P^{(1)}(x,a,T) = \sum_{t=1}^T \sum_y\sum_b P(y,b,t) P_{yb}(x,a,T-t),
\label{P1gen}
\end{equation}
where $P_{yb}(x,a,T-t)$ is the distribution obtained from a perfect
walk starting in state $\ket{y,b}$ for $T-t$ steps.
For the ideal walk, $\sigma^2(T) \equiv \sum_x\sum_a x^2 P(x,a,T)$,
and for the walk with decoherence,
\begin{equation}
\sigma^2(T,p) \equiv \sum_x\sum_a x^2 P(x,a,T,p).
\label{sddefgen}
\end{equation}
Taking equation~(\ref{eq:PxapTgen}) to first order in $p$,
and substituting along with equation~(\ref{P1gen}) into 
equation~(\ref{sddefgen}) gives to first order in $pT$
\begin{equation}
\sigma^2(T,p) \simeq \sum_{x,a} x^2 \left\{\mbox{\rule[-1em]{0em}{2.7em}}(1-pT)P(x,a,T) + p\sum_{t=1}^T \sum_{y,b} P(y,b,t) P_{yb}(x,a,T-t)\right\}.
\label{sigTp}
\end{equation}
The first term on the r.h.s. is (by definition) $(1-pT)\sigma^2(T)$.
Noting that $P_{yb}(x,a,T-t)$ is a translation of a walk starting at
the origin, $P_{yb}(x,a,T-t) = P_{0b}(x-y,a,T-t)$,
and relabelling the summed variable $x$ to $(x+y)$ then enables
the sums over $x$ and $a$ to be performed in the second term,
\begin{eqnarray}\label{sdpart}
&& p\sum_{t=1}^T \sum_y\sum_b P(y,b,t)\sum_x\sum_a (x+y)^2 P_{0b}(x,a,T-t) \nonumber\\
& = & p\sum_{t=1}^T \sum_y\sum_b P(y,b,t) \left\{\sigma^2_{0b}(T-t) + 2y\langle 
x\rangle_{0b}^{(T-t)} + y^2 \right\}.
\end{eqnarray}
From equation~(\ref{eq:sigma2}), $\sigma^2_{0b}(T-t)$ does not
depend on $b$, so the summation over $y$ and $b$ may be performed trivially.
The remaining summation over $y$ and $b$ applied to $y^2$ gives $\sigma^2(t)$ by definition.
This leaves only the evaluation of
\begin{equation}
 2p\sum_{t=1}^T \sum_{y,b}P(y,b,t)y\langle x\rangle_{0b}^{(T-t)}
 =  2p(1-1/\sqrt{2})\sum_{t=1}^T (T-t)\sum_{y,b}yb~ P(y,b,t),
\end{equation}
where we have used equation~(\ref{depmoments}) for $\langle x\rangle$.
We note that this term does not depend on whether
the initial coin state is plus or minus one and so we may
include both these possibilities equally.
Also, by the symmetry of the walk, it is possible to rewrite a 
probability function for travelling from state \ket{0,c} to \ket{y,b} in
the reverse order, i.e. as a probability for moving from \ket{y,b} to
\ket{0,c}.  Care must be taken to ensure that the signs of each term
due to the coefficient $yb$ in the summation are maintained.
We obtain
\begin{eqnarray}
&&2p\!\!\sum_{t,y,b}\!\!P(y,b,t)y\langle x\rangle_{0b}^{(T-t)}
= p(1-1/\sqrt{2})\sum_{t=1}^T (T-t)\!\!\sum_{y,b,c}(1-2\delta_{b,c})yb~P_{yb}(0,
c,t)\nonumber\\
&&= p(1-1/\sqrt{2})\sum_{t=1}^T
(T-t)\!\!\left[\sum_{y,b,c}yb~P_{y,b}(0,c,t)-2\sum_{y,b}yb~P_{yb}(0,b,t)]\right]
\!\!,
\end{eqnarray}
treating the two parts with and without a delta function independently.
Expanding the summations over $b$ and translating
the position basis by $-y$ gives
\begin{eqnarray}
&&2p\sum_{t=1}^T \sum_{y,b}P(y,b,t)y\langle x\rangle_{0b}^{(T-t)}\nonumber\\
& = &p(1-1/\sqrt{2})\!\!\sum_{t=1}^T(T-t)\!\!\left[\sum_{y,c}y P_{0,-1}(y,c,t)
-4\sum_y y~P_{y,+1}(y,+1,t)\right]\!\!\!.
\label{eq:term2}
\end{eqnarray}
The final summation over $y$ may be bounded above by noting that
\begin{eqnarray}
P_{0-1}(y,-1,t)&=&|\bra{y,-1}U^t\ket{0,-1}|^2\nonumber\\
               &=&\frac{1}{2}|\bra{y+1,1}U^{t-1}\ket{0,-1} -
               \bra{y+1,-1}U^{t-1}\ket{0,-1}|^2 \nonumber\\
               &\leq&\frac{1}{2}\sum_c P_{0,-1}(y+1,c,t-1).
\end{eqnarray}
Using this in equation~(\ref{eq:term2}) gives,
\begin{eqnarray}
2p\sum_{t,y,b}P(y,b,t)y\langle x\rangle_{0b}^{(T-t)}
& \leq & p(1-1/\sqrt{2})\sum_{t=1}^T(T-t) \left[\langle y
    \rangle_{0,-1}^t -2 \langle y \rangle_{0,-1}^{(t-1)}
    +2\right]\nonumber\\
&\leq&p(1-1/\sqrt{2})\sum_{t=1}^T(T-t)\left[(1-1/\sqrt{2})t+\sqrt{2}\right],
\label{eq:nearfin}
\end{eqnarray}
where equation~(\ref{depmoments}) has been used for the average values.
Note that equation~(\ref{depmoments}) is exact only for asymptotically large
times, so the second step in equation~(\ref{eq:nearfin}) introduces further
approximations from the contributions to the sum at small times $t$.
Combining these results in the full expression
for $\sigma^2(T,p)$, equations~(\ref{sigTp}, \ref{sdpart}), and
performing the summations over $t$ using
$\sum t= T(T+1)/2$ and $\sum t^2 = T^3/3 +T^2/2 +T/6$, gives
\begin{equation}
\sigma^2(T,p) \leq \sigma_0^2(T)\left[ 1 - \frac{\sqrt{2}}{6}pT +
p(\sqrt{2}-1)+ \ldots \right].
\end{equation}
Taking the square root gives as an upper bound on the standard deviation,
\begin{equation}
\sigma(T,p) \leq \sigma(T)\left[1-\frac{pT}{6\sqrt{2}} +
\frac{p}{\sqrt{2}}(1-1/\sqrt{2}) + O(p^2, 1/T)\right].
\label{eq:sigpTfin}
\end{equation}
Equation (\ref{eq:sigpTfin}) compares well with simulation data
in \cite{kendon02c}, once a second order correction
for $\sigma(T) = (1-1/\sqrt{2})^{1/2}(T-1/T)$ is taken into
account.  The bounding procedure
applied here is reasonably accurate, numerical studies give the
coefficient of $p$ in the above expansion as 0.09566, compared with the
bound of 0.20711.
The first order dependence is thus proportional to $pT$, 
the number of decoherence events during the whole quantum walk.
For a given decoherence rate $p$, 
the standard deviation initially decreases linearly in $T$.
This calculation first appeared in \cite{kendon02b}.


A similar expansion at the classical end of the full sum in
equation~(\ref{eq:PxapTgen}) gives
\begin{equation}
P(x,a,T,q) = (1-q)^TP^{(T)}(x,a,T) + q(1-q)^{T-1}P^{(T-1)}(x,a,T) + \dots ,
\label{eq:PxaqTgen}
\end{equation}
where we have defined $q\equiv(1-p)$ as the small parameter.
We know that $P^{(T)}(x,a,T)$ is the classical walk, so
\begin{equation}
\sigma^2_C(T) = T = \sum_x\sum_a x^2 P^{(T)}(x,a,T),
\end{equation}
and we can write
\begin{eqnarray}
\sigma^2(T,q) &=& \sum_x\sum_a x^2 P(x,a,T,q)\nonumber\\
              &=& (1-q)^T\sigma^2_c(T)
		+ q(1-q)^{T-1}\sum_x\sum_a x^2 P^{(T-1)}(x,a,T) + \dots .
\end{eqnarray}
However, to see a difference between the classical and quantum walks, we
need to have four consecutive steps of the quantum walk, since the
position distributions are identical for the first three steps.
The first term that differs from classical is thus derived from
part of $P^{(T-4)}(x,a,T)$ given by 
\begin{equation}
P^{(4cq)}(x,a,T) = \sum_{t=1}^{T-4}\sum_{y,z}\sum_{b,c}P^{(T)}(y,b,t)
	P_{(y,b)}^{(0)}(z,c,4)P_{y+z,c}^{(T)}(x,a,T-4-t),
\end{equation}
i.e., $t$ classical steps, 4 quantum steps, $T-4-t$ classical steps.
(Also included in $P^{(T-4)}(x,a,T)$ are combinations with the 4 quantum
steps not adjacent to each other.)
Since classical segments of the walk are not influenced by the initial
state, we may estimate the variance of the total contribution of
$P^{(4cq)}(x,a,T)$ by summing the variances of the segments:
\begin{equation}
\sigma^2_{(4cq)}(T) = \sum_{t=1}^{T-4} \left[\sigma^2_C(t) + \sigma^2(4)
	+ \sigma^2_C(T-t-4)\right]
	= (T-4)[T - 4 + \sigma^2(4)].
\end{equation}
Explicit calculation ($P(x,a,4)$ has sixteen terms) gives $\sigma^2(4) = 5$.
From the continued expansion of equation (\ref{eq:PxaqTgen})
the prefactor is $q^4(1-q)^{T-4}$,
so for small $q$ and large $T$, we have $\sigma^2(T,q) \simeq T(1+q^4)$
and for the standard deviation
\begin{equation}
\sigma(T,q) \simeq \sqrt{T}\left(1 + \frac{q^4}{2}\right),
\end{equation}
to lowest order in $q \equiv 1-p$.

A simpler model of decoherence in both position and coin space is
solved by \cite{romanelli04a}.  They apply measurements at regular
intervals and project the coin into the $\sigma_y$ basis, to preserve
the symmetry of the walk.  They obtain a spreading rate that is
basically classical, with a prefactor controlled by the rate of
the measurements.
The average squared distance from the start, measured by the variance,
grows as a series of arcs at the quantum rate (quadratic), with
regular resetting due to the measurements, pulling the overall
rate back to linear (classical).
They also generalise to random intervals between measurements,
drawing analogy with Brownian motion.

A unitary decoherence model affecting the position also appears in the
same paper \cite{romanelli04a} where they analyse a quantum walk in which
links between the positions on the line are broken with
probability $p$.  The transition that should have taken place
is turned into a self loop for that step of the walk.
The unitary operator for each step is modified to take account
of the links that happen to be broken at that time step.
For low rates of link-breaking, the quantum behaviour persists, 
while for high rates the walk makes less progress than even
an unimpeded classical random walk.

\subsection{Multiple coins in the walk on the line}
\label{ssec:manycoin}

Another analytically tractable approach to reducing coherences
in the walk on a line is to enlarge the size of the coin
state space, and use parts of it in turn as the walk progresses.
This has the advantage of the dynamics remaining purely unitary,
rendering the calculations simpler.
The first such study, by \cite{brun02b}, considered multiple
coins used in sequence, with the sequence repeating after all
the coins had been used once.  This produces a quantum walk
that still spreads linearly with the number of steps, but with the
rate of spreading reduced inversely by the number of coins.
Only if a new coin is used for every step of the walk does it
become equivalent to the classical random walk.  This is in 
contrast with the behaviour obtained by decohering the coin,
\citep{brun02c}, which always results in classical limiting behaviour,
as pointed out in \cite{brun02a}.
Classical behaviour is thus associated with an environment
so large that one never comes close to the Poincar\'e recurrence time
over the timescales considered.

Related studies that observe or exploit the behaviour when the
coin space is limited include \cite{flitney03a} who use
multiple coins to create a Parrondo game, by having the amplitude of the
coin flip depend on several previous coins rather than just one.
\cite{ermann05a} use an enlarged coin space to show explicitly that
the behaviour changes from classical back to quantum once the
size of the environment space is used up.  They use the coins in a random
order rather than sequentially as \cite{brun02c} do.
\cite{ribeiro04a} study quasi-periodic sequences of coin operators
using numerical simulation, and again find different spreading rates, but
still an overall linear dependence on the number of steps.
More realistic models of a finite-sized environment would have the
environment degrees of freedom interacting with each other, but this has
not yet been studied in the context of quantum walks.

\subsection{Effects in the walk on the $N$-cycle}
\label{ssec:cycdec}

Recall from \S\ref{ssec:2dcyc} that pure discrete-time quantum walks on cycles
don't mix (except instantaneously for a few small special cases mentioned
in \S\ref{ssec:pd}),
unless the time-averaged probability distribution, equation (\ref{eq:tvd})
is considered.  In that case, mixing to the uniform distribution
$\overline{M}(\epsilon)$, equation~(\ref{eq:meanmixdef}),
does occur for some choices of coin operator,
but this comes at the cost of requiring
a number of steps linear in the inverse accuracy $1/\epsilon$
instead of logarithmic, unless a number of repetitions are
combined in an amplification procedure.
\cite{kendon02c} carried out numerical studies of decoherence in cycles.
They evaluated $\overline{M}(\epsilon,p)$,
the mixing time for $\overline{P(x,T,p)}$
c.f.~equation~(\ref{eq:meanmixdef}),
for walks on cycles of sizes
up to $N\simeq 80$, both for pure states, and in the presence of the same
types of decoherence as described in \S\ref{ssec:lincoindec}
for the walk on a line.  They did not include any amplification procedure.
For odd-$N$ cycles with no decoherence,
they report that $\overline{M}(\epsilon) \sim N/\epsilon$ 
as compared to the upper bound of
$\overline{M}(\epsilon) \sim N \log N/\epsilon^3$
given by \cite{aharonov00a}.
\cite{richter06b} has recently confirmed this analytically.

\begin{figure}
            \begin{center}
            \resizebox{0.6\columnwidth}{!}{\includegraphics{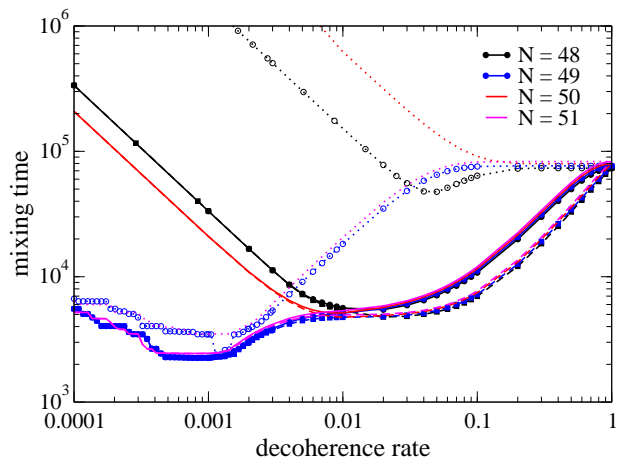}}
            \end{center}
            \caption{Numerical data for $\overline{M}_{\epsilon}(p)$
		on cycles of size $N=48$, $N=49$, $N=50$ and $N=51$,
		for coin (dotted), position
                (dashed) and both (solid) subject to decoherence,
                using $\epsilon=0.002$.  Both axes logarithmic.}
            \label{fig:c48c49c50c51mixav}
\end{figure}
Under the action of a small amount of decoherence, the mixing time becomes
shorter for all cases, typical results are shown in figure~\ref{fig:c48c49c50c51mixav}.
If the coin operator is chosen such that the even-$N$ cycles do not
mix to to the uniform distribution in the pure quantum walk,
then the addition of decoherence causes
them to mix to the uniform distribution.
Although for $N$ divisible by 4, the coin-decohered
mixing time shows a minimum below the
classical value at $p \simeq 2/N$, this mixing time is $\gtrsim N^2/32\epsilon$,
i.~e.,~still quadratic in $N$.
Thus, although noise on the coin causes the even-$N$ cycle to mix
to the uniform distribution, it does not produce a significant speed up over
the classical random walk.
Decoherence on the position produces a minimum mixing time
$\overline{M}_{\text{min}}(\epsilon,p) = O( N/\epsilon)$,
thus the even-$N$ cycle 
mixes to uniform in linear time for a suitable choice of decoherence
rate $p_{\text{min}} \sim 16/N^2$, independent of $\epsilon$.

For all types of decoherence, the odd-$N$ cycle shows a minimum mixing
time at a decoherence rate somewhat earlier than the even-$N$ cycle,
roughly $p = 2/N^2$,
but because of the oscillatory nature of $\overline{P(x,T,p)}$,
the exact behaviour is not a smooth function of $p$ or $\epsilon$.  
(\cite{kendon02c} illustrate this in their figure 4, we will show a similar
example later.)
As decoherence on the position (or both coin and position) increases, 
the oscillations in $\overline{P(x,T,p)}$ are damped out.
At $p \simeq 16/N^2$, the mixing time passes smoothly
through an inflexion and from then on behaves
in a quantitatively similar manner to the adjacent-sized even-$N$ cycles,
including scaling as
$\overline{M}_{\text{min}}(\epsilon,p) = O(N/\epsilon)$
at the inflexion.  Thus for $0 < p \lesssim 16/N^2$ there is a region
where the mixing time stays linear in $N$.

\begin{figure}
            \begin{center}
            \resizebox{0.6\columnwidth}{!}{\includegraphics{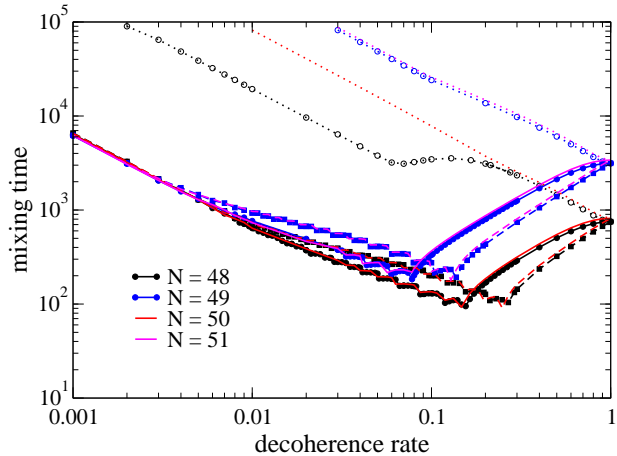}}
            \end{center}
            \caption{Numerical data for $M_{\epsilon}(p)$ on cycles of size
                $N=48$ to $N=51$,
		for decoherence applied to coin only (dotted),
		position only (dashed) and both (solid),
                using $\epsilon=0.002$.  Both axes logarithmic.}
            \label{fig:c48c49c50c51minst}
\end{figure}

However, one can do better than this: decoherence causes the quantum walk
to mix to the uniform distribution without any averaging of $P(x,T,p)$,
thus retaining the logarithmic scaling with
$\epsilon$ for the mixing times.  It is obvious that decoherence must do
this, since high decoherence rates reproduce the classical random walk,
which has this property.  The only question is whether it does so effectively
enough to be useful, and the answer is that it does, for decoherence
on the position \citep{kendon06c,richter06b}.
Figure \ref{fig:c48c49c50c51minst} shows the mixing time $M(\epsilon,p)$
corresponding to $P(x,T,p)$ as a function of $p$ for cycles of size 48 to 51.
Since even-sized cycles only have support on half the positions at any one time,
we have defined the uniform distribution only on the sites where it has
support (we can easily fix this if necessary by averaging over two consecutive
time steps).  It thus behaves like an odd-sized cycle of half the size.
Other than this technicality, the behaviour is the same, oscillating
peaks end abruptly at the minimum mixing time, followed by a smooth rise
to the classical value as the decoherence rate is turned up.
The decoherence rate at the minimum mixing time is
approximately $\pi/N$ (odd-sized) or $2\pi/N$ (even-sized), and the
mixing time itself scales as $O(N\log(1/\epsilon))$.  This thus provides a
quadratic improvement over the classical mixing time, which scales as
$O(N^2\log(1/\epsilon))$.  

\begin{figure}
            \begin{center}
            \resizebox{0.6\columnwidth}{!}{\includegraphics{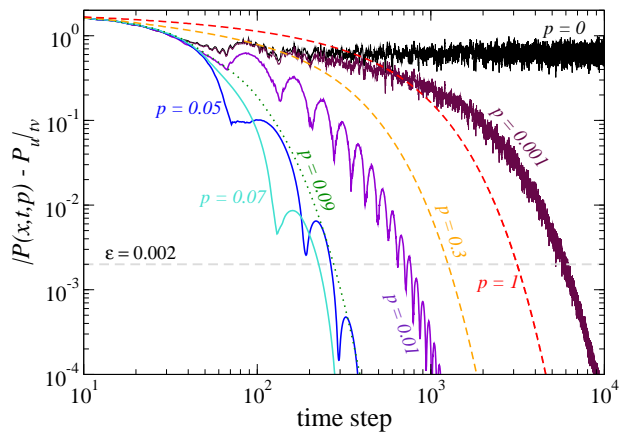}}
            \end{center}
            \caption{Difference between the probability
                distribution $P(x,T,p)$ and the uniform distribution
                expressed as the total variational distance,
                equation~(\ref{eq:tvd}), for $N=49$ for the case with
                decoherence on both coin and position.
                Both axes logarithmic.  The value of $\epsilon$ used in
                figures~\ref{fig:c48c49c50c51mixav} and
		\ref{fig:c48c49c50c51minst} is shown as a horizontal
		line.}
            \label{fig:c49b-epsilon}
\end{figure}

If we take a look at the behaviour of
$||P(x,t,p)-P_u||_{\text{tv}}$, shown in figure~\ref{fig:c49b-epsilon}
for $N=49$, we can see this too is fluctuating at the quantum end,
with the period of fluctuation lengthening as it reaches the minimum.
On the classical side of the minimum it increases smoothly to the classical
value.  The time at which the curves last cross the horizontal line
at $\epsilon=0.002$ is the mixing time as plotted in
figure~\ref{fig:c48c49c50c51minst}.
In the quantum regime, a different choice of
$\epsilon$ thus causes a jump in the value of
$M(\epsilon,p)$ if it happens to touch the next peak in
$||P(x,t,p)-P_u||_{\text{tv}}$.
This transition from underdamped (oscillating) to overdamped (smooth)
has been observed in a number of decoherence studies, such as
the hypercube \citep{kendon02c}, described in the next section,
and the studies of decoherence in continuous-time quantum walks
described in \S\ref{sec:qrwcontdec}.  The critical damping point
also seems to be associated with the point as which all
quantum correlations are destroyed by the decoherence
\citep{maloyer06a}, just as this marks the optimal top hat distribution
in the walk on the line.

\cite{richter06b} actually proves not only that the optimal mixing
time is $O(N\log(1/\epsilon))$, but also that three different strategies
for optimising the mixing time are equally effective.
The ``warm start'' from \cite{aharonov00a} can be viewed as running
the quantum walk for several segments of random length separated by
measurements.  The decoherence applied with probability $p$ is
a quantum walk of length $T$ interrupted by $pT$ measurements at
random times.  Simply running the quantum walk for several segments
of equal length with measurements applied at the end of each is also
effective: the random outcomes from the measurements are sufficient
to optimise the mixing.  This shows that the effects of decoherence
are quite robust, insensitive to the precise way in which it is applied.
The overall conclusion is thus the same as for the walk on a line, there
is a useful window within which decoherence enhances rather than degrades
the quantum features of the walk.  Moreover, it is actually necessary
to ensure the scaling with the precision $\epsilon$ is efficient.
Additionally, decoherence ensures the quantum walk mixes to a uniform
distribution, regardless of the initial conditions or choice of coin operator.

\subsection{Effects in the walk on the hypercube}
\label{ssec:hypdec}

\cite{kempe02a,kempe02c} picked up from where \cite{moore01a} left off and
analysed hitting times in discrete-time quantum walks on the hypercube.
By hitting time we mean the time it takes for the quantum walk
to reach the opposite corner from where it started from.
\citeauthor{kempe02a} considers two types of
hitting times, one-shot, where a measurement is made after a
predetermined number of steps, 
and concurrent, where the desired location is monitored
continuously to see if the walker has arrived.
While the one-shot hitting time is a pure quantum dynamics, the
continuous monitoring of the concurrent hitting time measurement 
removes part of the quantum coherences, 
thus it constitutes a type of decoherence.
More recently, 
\cite{krovi05a} have defined a hitting time in closer analogy with the usual
classical hitting time, the average time of first arrival at the 
target location.

Following \citeauthor{kempe02c}, the one-shot hitting time
$H_{\text{one-shot}}(r_0)$ between $\ket{x_0}$ and $\ket{x_f}$ is
\begin{equation}
H_{\text{one-shot}}(r_0) = \left\{T\enspace:\enspace|\bra{x_f} \mathbf{U}^T\ket{x_0}|^2 \ge r_0\right\},
\end{equation}
where $r_0$ is a threshold probability of being at the target state,
$\mathbf{U}$ is the evolution of one step of the walk, $\ket{x_0}$
is the starting state and $\ket{x_f}$ is the location being hit.

To define the remaining two hitting times, we need to define a measured
quantum walk, both \citeauthor{kempe02c}, and \citeauthor{krovi05a}
give essentially equivalent definitions.
For hitting a single final location we need a measurement with two 
outcomes (this is easily generalised), say $\mathbb{P}_f$ and
$\mathbb{Q}_f = \identity - \mathbb{P}_f$, where $\mathbb{P}_f = 
\ket{x_f}\bra{x_f}\otimes\identity_c$, i.e., the projector onto the
state $\ket{x_f}$ for any coin state.  If the walker is found at $\ket{x_f}$,
the walk is assumed to end, it acts as an absorbing boundary.
If the walk does not reach $\ket{x_f}$ after $t$ steps, the state of the
walk can be written
\begin{equation}
\bm\rho(t) = \frac{\mathcal{U}_Q^t\bm\rho}{Tr[\mathcal{U}_Q^t\bm\rho]}
\end{equation}
where $\mathcal{U}_Q\bm{\rho} = \mathbb{Q}_f\mathbf{U}\bm{\rho}\mathbf{U}^{\dagger}\mathbb{Q}_f^{\dagger}$.
This differs from equation (\ref{eq:Unonunitary}) in that we keep
only one outcome from the measurement and renormalise the density matrix.
Since we know the measurement outcome, we know whether the walk has
arrived or not, so we only need to keep
the part that did not yet arrive at $\ket{x_f}$.
The first arrival (also called first crossing) probability after $t$
steps can thus be written
\begin{equation}
r(t) = Tr\left[\mathcal{U}_P\mathcal{U}_Q^{t-1}\bm\rho\right].
\end{equation}
\citeauthor{kempe02c} defines the concurrent hitting time as
\begin{equation}
H_{\text{concurrent}}(r_0) = \min\left\{T\enspace |\enspace\exists\enspace t < T : r(t) > r_0\right\}
\label{eq:Hc}
\end{equation}
The hitting time defined by \citeauthor{krovi05a} is 
\begin{equation}
H_{\text{average}} = \sum_{t=0}^{t=\infty} t\, r(t),
\label{eq:Hq}
\end{equation}
the average arrival time.
\citeauthor{krovi05a} compare the concurrent hitting time with their average
hitting time and find they both scale in a similar polynomial manner,
compared with the exponentially long hitting time of a classical random
walk, provided they use a Grover coin for the quantum walk, and consider
$\ket{x_0} = \ket{0\dots 0}$ and $\ket{x_f} = \ket{1\dots 1}$.  Their 
numerical results suggest the bound of $H_{\text{concurrent}}(\Omega(1)) \le O(n^2\log n^2)$
obtained by \cite{kempe02a,kempe02c} is not tight, the actual hitting time
looks to be better than this in the large-$n$ limit.

\citeauthor{krovi05a} also investigate the effects of breaking the
symmetry of the Grover coin, by using a DFT coin, and by distorting the
hypercube lattice.  The DFT coin has the property that for certain
configurations the hitting time becomes infinite, while a
Grover coin with a
small distortion of the hypercube lattice
increases the hitting time somewhat, though it is still less
than the classical random walk hitting time.
Analogous results for coined quantum walks on the ``glued trees'' graph
were noted by \cite{tregenna03a}.
Further work by \cite{krovi06a} on the question of whether and
under what conditions quantum walks show dramatically different
properties (speed up or slow down) compared with classical random
walks suggests that it is highly dependent on the symmetry of the graph.
For the Grover coin, this is exemplified by the quantum walk search
algorithm of \cite{shenvi02a} (described in \S\ref{ssec:qwsearch}), where
any disturbance of the symmetry causes the walker to converge on
the marked state.

\cite{kendon02c} investigated the effects of adding
extra decoherence to the quantum walk on a hypercube over and above
that implied by the measurements of the target location.
For both one-shot and concurrent hitting times (this work
predates \cite{krovi05a}), the key parameter is the probability $r(t)$ of
finding the walker at the chosen location.
\begin{figure}
            \begin{center}
            \resizebox{0.6\columnwidth}{!}{\includegraphics{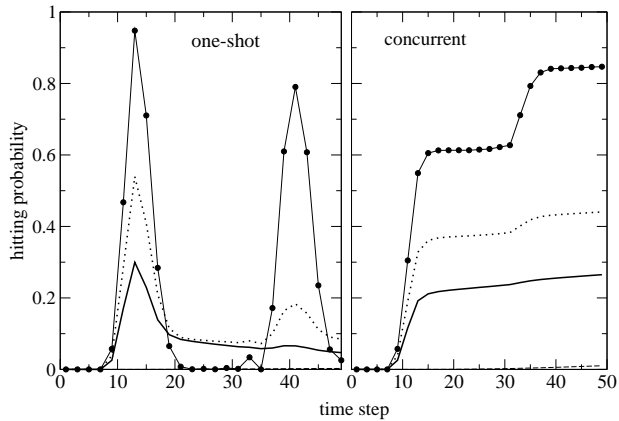}}
            \end{center}
            \caption{Hitting probability on a 9--dimensional hypercube
                for one-shot (left) and concurrent (right),
                perfect walk (circles),
                with $p = 0.05$ (dotted), $p = 0.1 \simeq 1/9$ (solid).
                Classical hitting probability barely visible (dashed).}
            \label{fig:9hit}
\end{figure}
Their numerical calculations show that
all forms of decoherence have a similar effect on $r(t)$,
see figure~\ref{fig:9hit}, reducing the peaks and smoothing out the troughs.  
For the one-shot hitting time this is useful, 
raising $r(t)$ in the trough to well above the classical value,
so it is no longer necessary to know exactly when to measure.
For $p \lesssim 1/n$, the height of the first peak scales as
$r_p(t) = r(t)(0) \exp\{-(n+\alpha)p\}$,
where $0\lesssim\alpha\lesssim 2$ depending on whether coin, position
or both are subject to decoherence.
An exponential decrease in the presence of decoherence sounds about as
bad as it could reasonably be, and for long times, of course, decoherence
reduces the walk to classical behaviour.
However, the hitting times are short, only $\sim n\pi/2$ steps,
so $p\simeq 1/n$ only lowers $r(t)$ by a factor of $1/e$.
For algorithmic purposes this is insignificant, only a factor of
order unity and thus still exponentially better than classical.
Note also that the
size of the graph (measured in number of nodes) is $2^n$,
so the decoherence has only a linear effect measured in terms of
the size of the graph.

The concurrent hitting time already includes a portion of decoherence,
no extra features are produced by the addition of unselective decoherence,
but there is still a range of $0 < p \lesssim 1/n$ within which
the quantum speed up is preserved.
Note that in both the one-shot and concurrent cases, $p\simeq 1/n$
is a critical damping rate, smoothing out the second peak (shown at
around 40 ($\equiv 3n\pi/2$) steps in figure~\ref{fig:9hit}).

Decoherence in discrete-time walks has thus provided us with a number
of notable common features, such as smoother  spreading, enhanced mixing, 
and a transition from under-damped (quantum) to over-damped (classical)
behaviour analogous to classical damping in a harmonic oscillator.
We will see in the next section that these features also appear for
the continuous-time quantum walk under decoherence.

\section{Decoherence in continuous-time walks}
\label{sec:qrwcontdec}

Less work has been done on decoherence in continuous-time quantum
walks, probably in part because the numerical simulations require
more resources (to integrate rather than iterate the dynamics).
Analytical calculations for the continuous-time walk are often simpler
than for the discrete-time walk due to the lack of a coin, but in
the case of decoherence this removes the simple case of decoherence
on the coin only that has been so successfully studied in the 
discrete-time walk.
Nonetheless, there are two notable recent analytical results.
\cite{fedichkin05a} obtain the scaling for mixing times for the
walk on the cycle for both small and large decoherence rates.
They augment this with numerical studies showing a minimum
mixing time for intermediate decoherence rates.
\cite{alagic05a} solve the decoherent walk on the hypercube for
all rates of decoherence.  These results will be discussed in
separate sections below.  First, we will set up a general
model for decoherence in continuous-time quantum walks analogous
to the basic model used for discrete-time walks.  Then we will
briefly examine the effects of decoherence in the continuous-time
quantum walk on the line, obtained numerically, to illustrate the
key features we can expect to see.

The pure state evolution of the continuous-time quantum walk
given by equation (\ref{eq:qcon}) can be expressed in density matrix
form as
\begin{equation}
\frac{d\bm\rho(t)}{dt} = -i\gamma\left[\mathbf{A},\bm\rho\right],
\end{equation}
where $[a,b] \equiv ab - ba$ denotes the commutator.
We can add a non-unitary decoherence to this in exact analogy to
the discrete-time method by writing 
\begin{equation}
\bm\rho(t+\delta t) = (1-p\delta t)\left(\bm\rho(t)-i\gamma \delta t\left[\mathbf{A},\bm\rho\right]\right)
+p\delta t\mathcal{P}\bm\rho,
\end{equation}
where $\mathcal{P}$ represents the action of the noise, for example,
a set of projectors $\{\mathbb{P}_j\}$, such that
$\mathcal{P}\bm\rho = \sum_j\mathbb{P}_j\bm\rho\mathbb{P}^{\dagger}_j$.
For uncorrelated noise, taking $\delta t \rightarrow 0$ this yields
\begin{equation}
\frac{d\bm\rho(t)}{dt} = -i\gamma[\mathbf{A},\bm\rho] - p\bm\rho +p\mathcal{P}\bm\rho,
\label{eq:qwcdec}
\end{equation}
The effect of the extra two terms is to reduce some or all of
the off-diagonal elements of $\bm\rho(t)$ at a rate $p$ per unit time.

\subsection{Effects in the walk on the line}
\label{ssec:clinedec}

\begin{figure}
\begin{center}
\resizebox{0.6\columnwidth}{!}{\includegraphics{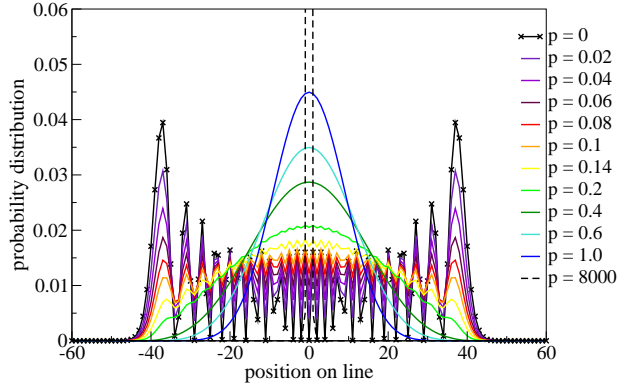}}
\end{center}
\caption{Probability distribution for a continuous-time quantum walk on the
	line with $\gamma=0.5$, at time $t=40$,
	for various decoherence rates $p$ give in the key.}
\label{fig:contdec}
\end{figure}

In contrast to the extensive studies of the discrete-time walk on the line,
there are no published analytical treatments of decoherence in a
continuous-time quantum walk on the line.  
We can easily investigate numerically:
for uncorrelated noise events at a rate $p$ per unit time,
we can unpack equation (\ref{eq:qwcdec}) into
\begin{equation}
\frac{d}{dt} \rho_{x,y}(t)
        = i \ \left[\frac{\rho _{x,y+1} - \rho_{x+1,y} - \rho_{x-1,y} + \rho_{x,
y-1}}{4}\right]
        - p \left({1 - \delta_{x,y}}\right)\rho_{x,y},
\label{eq:qwcdline}
\end{equation}
where $x,y\in\mathbb{Z}$.
(\cite{fedichkin05a} credit this model as developed by
\cite{gurvitz97a,gurvitz03a}.)
Figure \ref{fig:contdec} shows the results of evolving a continuous-time
quantum walk on the line with hopping rate $\gamma = 0.5$ until $t=40$,
for various decoherence rates.  This is to be compared with
figure \ref{fig:qvscwalk}, which shows the same type of study for
a discrete-time quantum walk.
The continuous-time walk does not produce quite such a nice
top hat as the discrete-time walk, partly due to having support on
all sites rather than alternate sites, though it would probably be
equally useful in practice.
There is, however, a striking new effect: unlike with the discrete-time
walk, where decoherence rates larger than one have no useful interpretation,
there is no reason why $p$ in equations (\ref{eq:qwcdec}) and
(\ref{eq:qwcdline}) cannot be made
arbitrarily large.  In figure \ref{fig:contdec}, $p=8000$ is shown: the
walk hardly manages to leave the starting point.  High values of $p$
can be interpreted as continuous monitoring, which projects the walk 
into the initial state with high probability.  In other words, the walk
suffers a quantum Zeno effect \citep{misra77a}, a phenomenon that 
does not appear in coined quantum walks.  One consequence of this is that 
quantities such as the concurrent hitting time are infinite,
since the continuous monitoring will ensure the
continuous-time quantum walk never arrives.

\subsection{Effects in the walk on the $N$-cycle}
\label{ssec:ccycdec}

The time-dependent non-unitary evolution of $\bm\rho(t)$ on a cycle
is also given by equation (\ref{eq:qwcdline}),
only with $x,y\in\mathbb{Z}^N$ for a cycle of size $N$.
 \cite{fedichkin05a} showed how to solve this
for small and large values of decoherence.
We will now outline their method.  The following substitution
\begin{equation}
R_{x,y} = i^{y-x} \rho_{x,y}
\end{equation}
converts the differential equations into a set with real coefficients,
\begin{equation}
\frac{d}{dt}{R_{x,y}} = \frac{1}{4}\left({R_{x,y+1} + R_{x+1,y} - R_{x-1,y} - R_{x,y-1}}\right) -
        p \left({1 - \delta_{x,y}}\right) R_{x,y}.
\label{eq:dRdt}
\end{equation}
Note that this leaves the diagonal elements of $\bm\rho$ identical to the
diagonal elements of $\bm R$.
As a side note, \cite{fedichkin05a} observe from these equations
that if $p=0$, there is an exact mapping of the quantum walk on a cycle
onto a classical random walk on a two-dimensional torus, and
if $p \ne 0$, there is still an exact mapping of the 
quantum walk on a cycle onto some classical dynamics on a directed toric graph. 
Similar mappings (one-dimensional quantum to two-dimensional classical)
also occur in quantum phase transitions in spin chains
\citep[see, for example,][]{sachdev99},
and can prove a useful technique for analysing quantum systems,
though we will not need to use it explicitly here.

Equation (\ref{eq:dRdt}) can be expressed as a linear operator equation
\begin{equation}
\frac{d}{dt}{\bm R(t)} = (\mathbb{L} + p\mathbb{U}) \ \bm R(t).
\label{eqn:operator}
\end{equation}
One can think of $\mathbb{L}$ and $\mathbb{U}$ as $N^2\times N^2$ matrices
and $\bm R(t)$ as a vector of size $N^2$ (instead of a matrix of size
$N\times N$). The indices of $\bm R(t)$ as a vector become $xN+y$ where
$x,y$ are the pair of indices of $\bm R(t)$ as a matrix, but we will
write $xy$ for such indices to reduce complexity in the notation.  
The formal solution of (\ref{eqn:operator}) can be
written $\bm R(t) = e^{t(\mathbb{L} + p\mathbb{U})}\bm R(0)$.
The explicit solution for arbitrary decoherence rates has not been obtained,
but one can examine the behaviour of the mixing time under the action
of small and large decoherence, using
standard perturbation theory methods.
When the decoherence rate $p$ is small such that $p N \ll 1$,
$p\mathbb{U}$ is treated as a perturbation of $\mathbb{L}$.
The unperturbed linear operator $\mathbb{L}$ (which evolves the pure state
quantum walk in density matrix formalism) has eigenvalues 
\begin{equation}
\lambda_{mn} 
        = i \ \sin\left(\frac{{\pi (m + n)}}{N}\right) \cos\left(\frac{{\pi (m -n)}}{N}\right)
\end{equation}
with corresponding eigenvectors
\begin{equation}\label{14}
V_{\mu\nu,mn} = \frac{1}{N} \exp\left(\frac{2\pi i}{N}(m\mu + n\nu)\right),
\end{equation}
Here $\mu\nu,mn$ are indices $0\le \mu,\nu,m,n < N$ combined like $x,y$ above.
With due consideration of the degeneracies in the eigenvalues 
$\lambda_{mn}$ of $\mathbb{L}$ \citep[see ][for details]{fedichkin05a},
the eigenvalue perturbation $\tilde{\lambda}_{mn}$ turns out to be
\begin{equation}
\tilde{\lambda}_{mn} = -p \frac{(N-2)}{N}.
\end{equation}
Thus, the solution is of the form
\begin{equation}
R_{xy}(t) 
        = \frac{\delta_{xy}}{N} + 
        \frac{1}{N^{2}} \sum_{mn} 
        (1 - \delta_{[m+n] (\mbox{\scriptsize mod }N),0}) \ 
        e^{t(\lambda_{mn} + \tilde{\lambda}_{mn})} \ 
        \exp\left[\frac{2\pi i}{N}(mx + ny)\right]
\end{equation}
The probability distribution of the continuous-time quantum walk is given
by the diagonal terms (considering $\bm R(t)$ as an $N\times N$ matrix again)
$P(x,t,p) = R_{x,x}(t)$, that is
\begin{eqnarray}
P(x,t,p)
        & = & \frac{1}{N} + 
                \frac{1}{N^{2}} \sum_{(m,n)} (1 - \delta_{[m+n](\mbox{\scriptsize mod }N),0}) 
                \times \left[\delta_{m,n} e^{-p \frac{N-1}{N}t} + (1-\delta_{m,n}) e^{-p \frac{N-2}{N}t}\right] \nonumber\\
        &   & \times \exp\left[it\sin\left(\frac{\pi(m+n)}{N}\right)\cos\left(\frac{\pi(m-n)}{N}\right)\right] 
                \exp\left[\frac{2\pi i}{N}(m + n)x\right] 
\label{eq:Pcdec}
\end{eqnarray}
\cite{fedichkin05a} then calculate an upper bound on the
mixing time $M(\epsilon)$.
Defining
\begin{equation}
\mathcal{M}_{x}(t) = \frac{1}{N} \sum_{m=0}^{N-1} e^{it\sin(2\pi m/N)} \omega_{N}^{mx},
\end{equation}
where $\omega_{N} = \exp(2\pi i/N)$,
and noting that 
\begin{equation}
\mathcal{M}_{x}^{2}(t/2) = \frac{1}{N^{2}} \sum_{m,n=0}^{N-1} e^{it\lambda_{(m,n)}} \omega
_{N}^{(m+n)x}, \ \ \
\mathcal{M}_{2x}(t) = \frac{1}{N} \sum_{m=0}^{N-1} e^{it\lambda_{(m,m)}} \omega_{N}^{2mx}
\end{equation}
and that $|\mathcal{M}_{x}(t)| \le 1$, we can simplify equation (\ref{eq:Pcdec})
as follows,
\begin{eqnarray}
\left|P(x,t,p) - \frac{1}{N}\right| 
        & \le & e^{-p \frac{N-2}{N}t} \left| \mathcal{M}_{x}^{2}(t/2) + \frac{e^{-tp/N} - 1}{N}
                \left[\mathcal{M}_{2x}(t) - \frac{2 - (N \mbox{ mod } 2)}{N}\right] \right
| \nonumber\\
        & \le & e^{-p \frac{N-2}{N}t} \ \left|1 + \frac{e^{-tp/N} - 1}
{N} (1 - 2/N)\right|.
\label{eq:Pinstmod}
\end{eqnarray}
Adding a now-trivial summation over $x$ gives us the total variational
distance required to obtain the mixing time:
\begin{equation}
\sum_{x=0}^{N-1} \left|P(x,t,p) - \frac{1}{N}\right| 
        \ \le \
        e^{-p \frac{N-2}{N}t} \ (N + e^{-tp/N} - 1).
\end{equation}
Since $e^{-tp/N} \le 1$, the above equation shows that the mixing condition is
$N e^{-p \frac{N-2}{N}t} \le \epsilon$.
This gives the mixing time bound of
\begin{equation}
M(\epsilon)
        \ < \ 
        \frac{1}{p} 
        \log\left(\frac{N}{\epsilon}\right)
        \left[1 + \frac{2}{N-2}\right].
\end{equation}
Some comments on this result are in order.  First of all, this is the 
mixing time for the instantaneous probability distribution, so it should 
be compared with the discrete-time results in
figure \ref{fig:c48c49c50c51minst}.  Just as for the discrete-time walk,
the first key effect of decoherence is to cause the continuous-time
quantum walk to mix.  On the face of it, $M(\epsilon)$ appears
to scale as $\log(N)$, which would imply an exponential speed up over
classical mixing times of $O(N^2)$.  However, this formula is only
valid for $pN\ll 1$, the $1/p$ scaling thus ensures that 
$M(\epsilon) \gg N$ for the range of validity of the result.
What this result tells us is that decoherence causes the continuous-time
quantum walk to mix with the dominant effect a scaling of $1/p$, and only
weak (logarithmic) dependence on $N$ and $\epsilon$.  It does not
tell us anything about the optimal decoherence rate that might give a minimum
mixing time, nor what value that minimum mixing time might take.

We can calculate the time-averaged mixing time from the
results for $P(x,t,p)$ above, and compare it with the discrete-time and 
classical values.  Again, we will obtain a result only valid in
restricted ranges of parameters.  Working with equation (\ref{eq:Pinstmod}),
and assuming $N\gg 1$ for simplicity, we obtain
\begin{equation}
\left|\overline{P(x,T,p)} -\frac{1}{N}\right| \le
\left|\frac{1}{T}\int_0^T\left(e^{-pt}+\frac{1}{N}\right)dt -\frac{1}{N}\right|
\le \left|\frac{1}{pT}(1-e^{-pT}) \right|
\end{equation}
Summing over $x$ to obtain the total variational distance trivially
multiplies by $N$.  Comparing this to $\epsilon$ as before gives
\begin{equation}
\frac{N}{pT}(1-e^{-pT}) \le \epsilon
\end{equation}
Assuming $pT\gg 1$ (which is consistent with our other assumptions,
$N \gg 1$ and $pN \ll 1$, it requires $T\gg N$) gives
\begin{equation}
\overline{M}(\epsilon) \ge \frac{N}{p\epsilon},
\end{equation}
which is similar to the scaling found numerically for the discrete-time
quantum walk.  In particular, the scaling with $N$ and $\epsilon$
is no longer logarithmic, as a consequence of the time-averaging.
The scaling with $p$ is still inverse, indicating that, in general,
pure continuous-time quantum walks on cycles do not mix to within
$\epsilon$ of uniform, even when time-averaged distributions are
considered, as already noted in \S\ref{ssec:qconcyc}.

We now turn to the analogous calculation for large decoherence rates.
When the decoherence rate $p$ is large,
that is, when $p \gg 1$, the quantum walk is experiencing a quantum
Zeno effect that suppresses most of the transitions.
We can therefore neglect all but the first off-diagonal terms
in the matrix $\bm R(t)$ from equation (\ref{eq:dRdt}). 
Retaining only matrix elements that are of order $1/p$ produces a
truncated set of differential equations for the elements along the major
and the two adjacent minor diagonals:
\begin{eqnarray}
\label{dRaa}
{R'_{x,x}}   & = & \frac{1}{4}\left(R_{x,x+1} + R_{x+1,x} - R_{x-1,x} - R_{x,x-1}\right), \\
\label{dRaa1}
{R'_{x,x+1}} & = & \frac{1}{4}\left(R_{x+1,x+1} - R_{x,x}\right) - p R_{x,x+1}, \\
\label{dRa1a}
{R'_{x,x-1}} & = & \frac{1}{4}\left(R_{x,x} - R_{x-1,x-1}\right) - p R_{x,x-1}.
\end{eqnarray}
To facilitate our subsequent analysis, we define
\begin{equation}
a_x = R_{x,x}, \ \ \ \ \ d_x = R_{x,x+1} + R_{x+1,x}.
\end{equation}
Then, we observe that
\begin{equation}
a'_{x} = \frac{\left(d_x - d_{x-1} \right)}{4}, \ \ \ \ \
d'_{x} = \frac{\left(a_{x+1} - a_x\right)}{2} - p d_x.
\end{equation}
The solution of differential equation at large $p$ has the form
\begin{equation}
a_x = \frac{1}{N}\sum_{y=0}^{N-1} \
        \exp{\left(-\frac{\sin^2{\frac{\pi y}{N}}}{2p} t\right)} \ \omega^{xy}.
\end{equation}
Based on the above analysis, the full solution for $R(t)$ is given by
\begin{equation} \label{eqn:large-solution}
R_{x,y}(t) =
        \left\{\begin{array}{ll}
        a_{x}           & \mbox{ if ~$x = y$ } \\
        d_{x}/2         & \mbox{ if ~$|x - y| = 1$ } \\
        0               & \mbox{ otherwise }
        \end{array}\right.
\end{equation}

The total variation distance between the uniform distribution and the probability distribution of the
decoherent quantum walk is given by
\begin{equation} 
\sum_{x=0}^{N-1} \left| {a_x(t) - \frac{1}{N}} \right| = 
        \sum_{x=0}^{N-1} \left|\frac{1}{N} \sum_{y=0}^{N-1} 
                \exp\left(-\frac{\sin^2{\frac{\pi y}{N}}}{2p} t\right)
                \exp\left(\frac{2\pi ixy}{N}\right) - \frac{1}{N}\right|,
\end{equation}
which simplifies to
\begin{equation} 
\sum_{x=0}^{N-1} \left| {a_x(t) - \frac{1}{N}} \right| =
      \frac{1}{N} \sum_{x=0}^{N-1} \left| \sum_{y=1}^{N-1}
        \exp\left(-\frac{\sin^2{\frac{\pi y}{N}}}{2p} t\right)
        \cos{\left(\frac{2\pi y x}{N}\right)} \right|.
\end{equation}

A lower bound on the mixing time for large decoherence rate
$p$ can be derived as follows.  Note that
\begin{eqnarray}
\sum_{x=0}^{N-1} {\left| {a_x(t) - \frac{1}{N}} \right|}
        & \ge & {\left| {a_0(t) - \frac{1}{N}} \right|}
        =   \frac{1}{N} \sum_{y=1}^{N-1} \exp{\left(-\frac{\sin^2{\frac{\pi y}{N}}}{2p} t\right)}, \\
        & \ge & \frac{2}{N} \exp{\left(-\frac{\sin^2{\frac{\pi}{N}}}{2p} t\right)},
\end{eqnarray}
where the first inequality uses the term $x = 0$ only and the second inequality 
uses the terms $y = 1,N-1$.
This expression is monotone in $t$, and is a lower bound on the total variation 
distance.  It reaches $\epsilon$ at time $T_{lower}$, when
\begin{equation}
T_{lower} 
        \ = \ 
        \frac{2p}{\sin^{2}{\frac{\pi}{N}}} \ln\left( \frac{2}{N \epsilon
} \right)
        \ \simeq \ 
        \frac{2p N^2}{\pi^2} \ln\left(\frac{2}{N\epsilon}\right),
\end{equation}
for large $N \gg 1$.

An upper bound on the mixing time for large decoherence rate $p$ can also
be derived (details in \cite{fedichkin05a})
\begin{equation}
T_{upper} \ = \ \frac{p N^2}{2} \ln\left(\frac{2+\epsilon}{\epsilon}\right).
\end{equation}

For large decoherence rates $p \gg 1$, the mixing times are bounded as 
\begin{equation}
\frac{p N^2}{\pi^2} \ln\left(\frac{2}{N \epsilon}\right) 
\ < \ M(\epsilon) \ < \
\frac{p N^2}{2} \ln \left(\frac{2+\epsilon}{\epsilon}\right).
\end{equation}
These bounds show that $M(\epsilon)$ is linearly proportional to the decoherence
 rate $p$,
but is quadratically dependent on $N$, and logarithmically dependent on
$\epsilon$.

The change in the form of the scaling with $p$ from reciprocal for
small $p$ to linear for large $p$ suggests there is a minimum
mixing time at some intermediate decoherence rate.
\begin{figure}
\begin{center}
\resizebox{0.6\columnwidth}{!}{\includegraphics{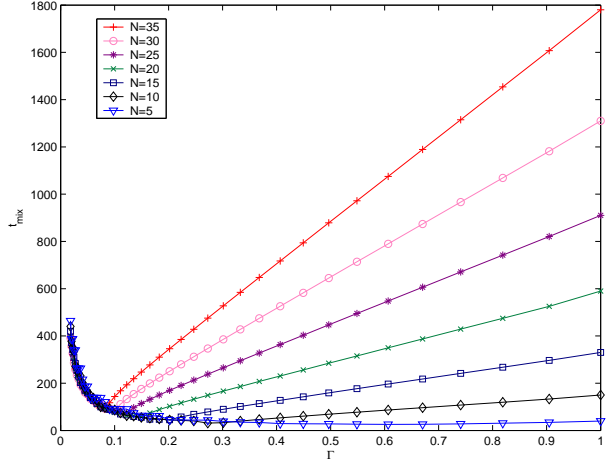}}
\end{center}
\caption{
The quantum to classical transition of mixing time in a continuous-time
decoherent quantum walk on $C_{N}$, for $N=5,10,15,20,25,30,35$.
(From \cite{fedichkin05a}.)
}
\label{fig:dima}
\end{figure}
\cite{fedichkin05a} confirmed this numerically,
reproduced in figure \ref{fig:dima}.

The effects of decoherence on continuous-time quantum walks on cycles is thus
similar to the effects on discrete-time walks on cycles, as described in
\S\ref{ssec:cycdec}.
In fact, the proofs in \cite{richter06b} are sufficiently general they apply
equally to the continuous-time walk as to the discrete-time walk.
This fills in the gap between small and large decoherence rates,
and proves that continuous-time walks, with $O(\log(1/\epsilon))$
decoherence or measurement events, mix in time $O(N\log(1/\epsilon))$
on the cycle (and the $d$-dimensional torus).

\subsection{Effects in the walk on the hypercube}
\label{ssec:chypdec}

The hypercube is an attractive subject for analytical
calculations of quantum walks because, under certain
conditions, it offers two routes to simplify the analysis.
If the walk has the appropriate symmetry, it is constrained
to a subspace that maps to a walk on a line segment
with positions corresponding to the Hamming weight of the
node labels (with position dependent transition rates).
Alternatively, the continuous-time walk can be factored in Fourier space
into independent walks on each qubit of the labels of
the locations, as explained in \S\ref{ssec:hypercube}.
Furthermore, both mixing properties for spreading over the
whole graph, and hitting properties for traversing to opposite
corners are non-trivial on the hypercube.
We thus have a large number of comparisons to make between
discrete and continuous-time quantum walks, and classical
random walks.
\cite{alagic05a} provide a complete solution to the dynamics
of the continuous-time quantum walk on the hypercube subject
to decoherence.  We will sketch their method of solution,
then discuss their results.

\citeauthor{alagic05a} give their
quantum walk an energy $k$ -- this is equivalent an arbitrary
hopping rate $\gamma = k/n$ instead of the choice of $\gamma=1/n$
for a graph with vertices of degree $n$.
We consider the continuous quantum walk on the $n$-dimensional hypercube
with energy $k$ and decoherence rate $p$, starting from the initial
wave function $\Psi_0 = \vert 0 \rangle ^{\otimes n}$, corresponding 
to the corner with Hamming weight zero. 
The decoherence operators in equation (\ref{eq:qwcdec}) project the
walker onto one of the vertices of the hypercube chosen uniformly at
random.  It is useful to write these projectors explicitly in terms of
single qubit projectors $\Pi_0$ and $\Pi_1$ onto $\vert 0 \rangle$
and $\vert 1 \rangle$, respectively. We have
\begin{equation} 
\mathbb{P} = \frac{1}{n} \sum_{1 \leq j \leq n} [\Pi^j_0 \otimes \Pi^j_0 + \Pi^j_1 \otimes \Pi^j_1] 
\end{equation}
where $\Pi^j_0 = \identity \otimes \cdots \otimes \identity \otimes \Pi_0 \otimes 
\identity \otimes \cdots \otimes \identity$ with the non-identity projector 
appearing in the $j$th place and similarly for $\Pi^j_1$.

We now show that, with the model of
decoherence described above, each dimension still behaves
independently.   Recall equation (\ref{eq:hypA}) showing that the
adjacency matrix for the hypercube decomposes into a sum of tensor products
each acting only on a single qubit.  Since the noise operators
$\mathbb{P}_j$ also have this structure, provided $\bm\rho(0)$ starts off
in a state that is also decomposable (such as the all-zero vertex),
the subsequent evolution will maintain this structure and will remain
decomposable as a system of $n$ non-interacting qubits.
Since the qubits are now in mixed states $\bm\rho_j$, the superoperator
acting on them is equivalent to a $4\times 4$ matrix
\begin{eqnarray}
 \mathcal{J} 
&=& \frac{t}{n}[(\identity \otimes in\sigma_x) - (in\sigma_x \otimes \identity) 
- p({\identity} \otimes {\identity})
   + p(\Pi_1 \otimes \Pi_1) + p(\Pi_0 \otimes \Pi_0)] \nonumber\\
&=& \frac{t}{n}\left( \begin{matrix} 0   &  ik & -ik &  0 \\
                        ik  & -p  &  0  & -ik \\
                        -ik &  0  & -p  &  ik \\
                        0   & -ik &  ik &  0 \\
    \end{matrix} \right).
\end{eqnarray}
Notice that for $p = 0$, $\left[e^{\mathcal{J}}\right]^{\otimes n}
= \left[e^{-it\sigma_x} \otimes e^{it\sigma_x}\right]^{\otimes n}$, 
which is exactly the superoperator formulation of the dynamics of 
the non-decohering walk. 
So far we have shown that the walk with decoherence is still
equivalent to $n$ non-interacting single-qubit systems. We now analyse
the behaviour of a single-qubit system under the superoperator
$e^\mathcal{J}$.
The structure of this single particle walk will allow us to then
immediately draw conclusions about the entire system. 

The eigenvalues of $\mathcal{J}$ are $0$, $- \frac{pt}{n}$,
$\frac{-p t - i\beta t}{2n}$ and $\frac{-p t +i\beta t}{2n}$.
Here $\beta = \sqrt{16k^2 - p^2}$ is a complex constant that will later turn out to be
important in determining the behaviour of the system as a function of
the rate of decoherence $p$ and the energy $k$. The matrix exponential
of $\mathcal{J}$ in this spectral basis can be computed by
inspection. To see how our superoperator acts on a density matrix
$\bm\rho_0$, we may change $\bm\rho_0$ to the spectral basis, apply the diagonal
superoperator to yield $\bm\rho(t)$, and finally change
$\bm\rho(t)$ back to the computational basis.  At that point we can apply the
usual projectors $\Pi_0$ and $\Pi_1$ to determine the probabilities of
measuring $0$ or $1$ as a function of time.
The result \citep[details in][]{alagic05a} is
\begin{equation*}
P[0] = \frac{1}{2} + \frac{1}{2} e^{-\frac{pt}{2n}} \left[\cos\left(\frac{\beta 
t}{2n}\right) + 
                         \frac{p}{\beta}~\sin\left(\frac{\beta t}{2n}\right) \right] 
\end{equation*}
\begin{equation}
P[1] = \frac{1}{2} - \frac{1}{2} e^{-\frac{pt}{2n}} \left[\cos\left(\frac{\beta 
t}{2n}\right) + 
                         \frac{p}{\beta}~\sin\left(\frac{\beta t}{2n}\right) \right].
\label{eq:Phyp}
\end{equation}
All the properties of the decohering walk can be deduced from these equations.
Just as was observed in the mixing behaviour of cycles,
the decoherence exhibits three distinctive regimes analogous
to underdamping, critical damping and
overdamping in a damped oscillator.  We will now describe the
properties of these regimes.

The underdamping regime occurs for $p < 4k$.
To determine the mixing times, we solve $P[0] = P[1] = \frac{1}{2}$,
giving exact instantaneous mixing times at
\begin{equation}
  M_{\text{inst}} = \frac {n (2\pi c - \arccos(p^2/8k^2-1))}{\sqrt{16k^2 -  p^2}} 
\end{equation}
for all $c \in \mathbb{Z}$, $c > 0$.  At these times, the total variational
distance between the walk distribution and the uniform distribution is zero,
but the walk ``unmixes'' again as time progresses.
This corresponds to the result in~\cite{moore01a} for pure state quantum
walks, and extends it to regions with
small decoherence rates.  These mixing times scaling linearly
with $n$ represent a quantum improvement
over the classical random walk mixing time of
$\Theta(n \log n)$, with the caveat that the classical mixing time is not
instantaneous, so strictly speaking we are not comparing like with like, in
the quantum case one must sample at \textit{exactly} the right moment in time.
Note that these periodic mixing times decay with $p$ and
disappear altogether when $p \geq 4k$, so the quantum advantage tails off
before the critical damping point is reached.
\cite{richter06b} offers a weak bound of $O(n^{3/2}\log(1/\epsilon))$
for the mixing time at around the critical damping point, which is slower
than classical.

By examining the local maxima of $P[1]$, we can determine that
the walk has approximate instantaneous hitting times to 
the opposite corner $(1, \dots , 1)$ at times
\begin{equation}
H_{\text{one-shot}}(r) = \frac{2 \pi n (2c + 1)}{\sqrt{16k^2 - p^2}}
\end{equation}
for all $c \in \mathbb{Z}$, $c \geq 0$. However, the probability $r(p,c)$ of
measuring an exact hit decays exponentially,
\begin{equation}
r(p,c) = \left[\frac{1}{2} + \frac{1}{2}e^{-\frac{p \pi (2c + 1)}
    {\sqrt{16k^2 -  p^2}}}\right]^n\enspace.
\end{equation}
When no decoherence is present, the walk hits at 
$H_{\text{one-shot}}(1) = \frac{n \pi(2c+1)}{2k},$ and it does so exactly, i.e. 
$r(c) = 1$.  This corresponds to the result of \cite{kempe02a,kempe02c}
for the discrete-time quantum walk, see \S\ref{ssec:hypdec}, thus both
discrete-time and continuous-time quantum walks on the hypercube exhibit
exponentially faster hitting times than classical random walks.
The difference is that the continuous-time quantum walk hits exactly, while
the discrete-time quantum walk hits with a probability that is less
than one, see \cite{kempe02a,kempe02c} for details.

For a fixed $p \geq 4k$, \cite{alagic05a} show that the walk
behaves much like the classical walk on the hypercube, 
the measurement distribution of the walk 
converges to the uniform distribution in time $M(\epsilon) = \Theta(n \log n)$,
just as in the classical case.

As $p \to \infty$, the walk suffers from
the quantum Zeno effect. Informally stated, the rate of decoherence is
so large that the walk is continuously being reset to the
initial wave function $|0\rangle^{\otimes n}$ by measurement.
By inspection of equations (\ref{eq:Phyp}), it is clear that
$P[0] \to 1$ and $P[1] \to 0$ as $p \to \infty$.

This concludes our overview of decoherence in continuous-time walks.
We have observed some of the same effects of enhanced mixing and
smoother spreading, and also shown that where the decoherence doesn't
directly help, the quantum speed up is at least robust under small
amounts of decoherence.

\section{Quantum walks in physical systems}
\label{sec:qrwphys}

As well as the potential for algorithmic applications that has been
in the background of most of our discussion of decoherence in quantum
walks in this review, physical systems can be made to perform a quantum
walk, and this process is of interest in its own right as an example
of precise coherent control.  Indeed, the first suggested application
of a quantum walk in a physical system was to test decoherence
rates in a single trapped ion.  \cite{travaglione01a} describe how
a coined quantum walk can be performed using the vibration modes of
the trapped ion coupled to an internal state of the ion for the quantum coin.
They note that a cycle of size $N=4$ performed in this way should exhibit
periodic behaviour, see \S\ref{ssec:pd}.
\cite{sanders02a} suggest a quantum walk using the phase of the 
electromagnetic field in a cavity as the position (so the walk is
again on a cycle) with the role of the coin played by an atom,
also in the cavity, which couples to the cavity field depending on
its internal state.  Since the cavity field cannot be in a state
localised at a point in phase space, the positions must be approximated
by small regions that do not overlap.  \citeauthor{sanders02a} provide
a detailed analysis of the experimental parameters necessary to maintain
distinct positions during the quantum walk.  
\cite{dur02a} propose a quantum walk on the line using 
an atom hopping between traps in an optical lattice.  The role of the
coin is again played by the internal state of the atom.  
One of the potential sources of error in this system is imperfect
transitions between lattice sites, so, along with analysing errors
due to imperfect Hadamard operations, \citeauthor{dur02a} also
consider the effects of imperfect shift operations.  Their work
is numerical, but shows that imperfect shifts, or spontaneous tunnelling,
smear out the distribution without necessarily reducing the quantum features.
This analysis is typical of the extra factors that need to be considered
in experimental systems.  Since the quantum walk takes place in a subspace
of the full quantum system, many of the potential sources
of decoherence take the system outside of the subspace of the quantum walk,
so the resulting decohered dynamics is not a classical random walk.
All of these proposals are within current experimental
capabilities but, to date, none has been carried out.

The above examples are all of discrete-time coined quantum walks.
\cite{solenov05a,solenov05b} analyse a ring of quantum dots in which
the electrons in the dots perform a continuous-time quantum walk.  
They apply similar perturbation theory techniques to analyse the
effects of decoherence as were described in \S\ref{ssec:ccycdec}.

There are a number of ways in which photons can be made to follow a
quantum walk dynamics, and while some have been implemented, all
have used high light intensities that admit a classical wave description.
The earliest example, which predates the algorithmic discussions of
quantum walks, is from \cite{bouwmeester99a}, who present the
twin peaked distribution of a quantum walk on the line without 
identifying it as such.  \cite{knight03a} discuss this experiment, 
arguing that quantum walks on the line are actually not
quantum at all.  \cite{kendon04a} explain how
what \cite{bouwmeester99a} have done can be regarded as a quantum walk 
and suggest an enhanced experiment, in which monitoring the individual photon
``walkers'' reduces the dynamics to a classical random
walk, to distinguish a classical ``wave walk'' from a quantum walk.

A quantum walk experiment carried out by
\cite{ryan05a} uses a 3 qubit NMR system to perform 
a quantum walk on a cycle of size $N=4$.  This is actually a quantum
computation of a quantum walk, since the three qubits are used to
represent the binary number labelling the vertex (two qubits),
and the qubit coin.  This is distinctly different from the physical
systems described above, see \cite{kendon06a} for more discussion
on this point.  \citeauthor{ryan05a} also apply decoherence artificially
to the qubits, to observe that the quantum walk degrades into the
classical walk.

\section{Quantum walks on directed graphs}
\label{sec:directed}

Providing a quantum walk dynamics on undirected graphs is straightforward
since the adjacency matrix of an undirected graph is symmetric, hence it 
can be turned into a unitary (with added coin) or Hermitian operator for
any such graph.
In contrast, determining whether a directed graph allows a unitary
dynamics that respects the graph structure is a hard problem, to which
a general answer is not known.  \cite{aharonov00a} give a useful definition
of how a quantum walk dynamics should respect the structure of a graph, though 
other approaches are possible, see, for example, \cite{montanaro05a}.
\cite{severini02a,severini03a} provides
some necessary conditions, by studying the pattern of a unitary matrix,
defined as $U^{(P)}_{ij} = 1$ if $U_{ij} \ne 0$ and $U^{(P)}_{ij} = 0$
if $U_{ij} = 0$.  If $A = U^{(P)}$ for some $U$, then a quantum walk can
naturally be applied to the graph with adjacency matrix $A$.
The situation is simpler for some restricted types of graphs, such as
line graphs.  A line graph $L^{(G)}$ is obtained from graph $G$
by making each edge in $G$ a vertex in $L^{(G)}$, and edges for each
pair of edges in $G$ that are adjacent.  If $x$, $y$, $z$ are vertices in
$G$ connected by directed edges $(xy)$ and $(yz)$ which can be traversed only
from $x$ to $y$ to $z$, then the vertices in $L^{(G)}$ correspond to all
$(xy) \in G$, and the edges in $L^{(G)}$ correspond to all
$((xy)(yz))$ where $(xy)$ and $(yz)$ are edges in $G$.
This definition can be applied recursively to generate a family of
line graphs from a single initial graph $G$, and \cite{pakonski02a}
provide conditions under which line graphs can be quantised.

Starting with an undirected graph, a corresponding directed graph
can be constructed by a mapping due to \cite{watrous98a}, similar to the
line graph definition above, only each undirected edge is treated as a
pair of directed edges.  The resulting directed graph is always the pattern
of a unitary matrix.  The corresponding dynamics,
using an associated unitary matrix as the step operator, 
corresponds exactly to the coined quantum walk dynamics (with a
Grover coin) on the original
undirected graph \citep{watrous02a}.

Directed graphs that are not the pattern of a unitary matrix may still
admit a quantum walk dynamics. \cite{montanaro05a} defines
``reversibility'' for a directed graph, and proves this allows 
for a unitary quantum walk in which the coin
selects between different sets of cycles in the graph, rather than
different paths from each vertex.  For irreversible graphs,
\citeauthor{montanaro05a} suggests splitting the graph into subgraphs that
are reversible, on which the quantum walk dynamics can be unitary,
combined with a non-unitary step (a partial measurement) 
to switch between subgraphs in an irreversible way that nonetheless
preserves coherence within each reversible subgraph.
If one is willing to allow some transitions that are between vertices
not connected by edges, the method of \cite{szegedy04a} also
provides a quantum Markov process on directed graphs.

This brief section only touches on the range of problems in the realm of
directed graphs, reflecting the limited work done in this area.
For example, no study of decoherence in quantum walks on
directed graphs has been done, nor consideration of the class of
graphs that might be the pattern of a CP map (rather than of a unitary
matrix).

\section{Summary and outlook}
\label{sec:conc}

We have seen in our broad-ranging review of decoherence in quantum walks
that the effects are both non-trivial and have potential algorithmic
applications.  As the decoherence rate is increased,
for moderate timescales the quantum features persist,
usually to a point of ``critical damping'' after 
which the behaviour resembles that of a classical system.
In the quantum region before the critical damping point, the 
useful algorithmic properties, such as spreading and mixing,
are often better than the pure quantum dynamics.
Indeed, for cycles and tori, some element of non-unitary evolution
is essential for efficient mixing behaviour.
Our tentative conclusion is thus that decoherence is a useful
enhancement, and non-unitary dynamics are a promising avenue of
enquiry in the quest for practical applications for quantum computing.
The incomplete nature of our conclusions are in part a reflection of
our limited knowledge of the conditions under which a quantum speed
up can be obtained.  

The idea that the addition of classical randomness can enhance a useful
algorithmic property also occurs in other settings, both quantum and classical.
An example from the field of quantum information may be found in
\cite{kraus04a,renner05a}, where the security of a quantum secret key
is increased by adding a small amount of randomness to the key
during the privacy amplification stage of the protocol.  This can be
understood in two ways.  The goal is to reduce the amount of information
available to an eavesdropper to a trivial level.  The added randomness
reduces the final size of the secret key, but it reduces even more
the information that the eavesdropper has.  This works for both classical 
and quantum keys.  From a quantum perspective,
the randomness takes the form of entanglement with an environment
that the eavesdropper does not have access to.  Since the total amount
of entanglement is limited, this reduces the entanglement of the key
with the eavesdropper, and thus reduces the information available to
the eavesdropper. 
The interpretation of this in the context of quantum walks is of course 
rather different, but the common theme is that the decoherence or
randomness induces a more uniform distribution, in this case in the
form of a smoother ``top-hat'' distribution in the walk on the line,
and faster mixing times on finite graphs.  Some further insight is
provided in \cite{kendon06c}, where the entanglement between the
coin and the position is used to gauge the impact of the added randomness.
The optimal decoherence rate turns out to just remove all the quantum
entanglement, and can be thought of as removing the ``extra'' quantum
correlations that would otherwise keep the distribution away from uniform.

As will have become clear to the reader by now, there are many gaps in
the analysis of decoherence in quantum walks, and, more importantly,
many gaps in our understanding of what might make a useful quantum
walk algorithm.  Some of these shortcomings will only be accessible
to serious study when (if) we have a larger (that is, larger than we
can simulate classically) quantum computer available to test the
performance of actual algorithms.  In the meantime, there is still
much useful progress that can be made, in particular in understanding
the role of symmetry in the properties of quantum walks.  Many of
the examples studied so far turn out on closer inspection to be 
exceptional cases, a slight change in the symmetry of the dynamics
or initial state can reverse the interesting quantum properties.
The role of decoherence as a ``memory eraser'' to eliminate the
dependence on the initial state, while still retaining the 
interesting quantum properties, is thus one of its most appealing
features.  My hope in offering this review is that it
will be a useful summary of the current state of research in a field
that has seen significant recent activity and advances, and
that it will facilitate and spur further efforts in this area.

I gratefully thank my colleagues for many useful and stimulating discussions of quantum walks, especially
Dorit Aharonov,
Andris Ambainis,
Todd Brun,
Ivens Carneiro\footnote{It is with much sadness we learned of the death of Ivens
Carneiro in a traffic accident in April 2006.},
Hilary Carteret,
Andrew Childs,
Richard Cleve,
Jochen Endrejat,
Ed Farhi,
Will Flanagan,
Mathieu Girerd,
Mark Hillery,
Peter H\o{}yer,
Julia Kempe,
Peter Knight,
Norio Konno,
Barbara Kraus,
Meng Loo,
Rik Maile,
Olivier Maloyer,
Cris Moore,
Peter Richter,
Eugenio Rold\'{a}n,
Alex Russell,
Barry Sanders,
Mario Szegedy,
John Sipe,
Tino Tamon,
Ben Tregenna,
John Watrous,
and
Xibai Xu.


\bibliography{../bibs/ent,../bibs/qit,../bibs/qrw,../bibs/misc,../bibs/shor}



\end{document}